\documentclass[amsmath,twocolumn,prb,aps]{revtex4-1}
\usepackage{amsthm,amsfonts,graphicx,verbatim, color}
\usepackage{bm}
\usepackage[T1]{fontenc}
\usepackage[utf8]{inputenc}

\newcommand{\aE}{\ensuremath{\alpha}}
\newcommand{\estimate}[1]{A(#1)}
\newcommand{\x}{\ensuremath{\gamma}}
\newcommand{\y}{\ensuremath{\sigma}}
\newcommand{\D}{\ensuremath{D}}
\newcommand{\ev}[1]{\left \langle #1 \right \rangle}
\newcommand{\eq}[1]{Eq.~\eqref{eq:#1}}

\newcommand{\mb}[1]{\ensuremath{\mathbf #1}}
\newcommand{\mc}[1]{\ensuremath{\mathcal #1}}
\newcommand{\bs}[1]{\ensuremath{\boldsymbol #1}}

\begin{document}

\title{Geometry of flux attachment in anisotropic fractional quantum Hall states}

\author{Matteo Ippoliti${}^1$, R. N. Bhatt${}^2$, and F. D. M. Haldane${}^1$}
\affiliation{${}^1$Department of Physics and ${}^2$Department of Electrical Engineering, Princeton University, Princeton NJ 08544, USA}

\begin{abstract}
Fractional quantum Hall (FQH) states are known to possess an internal metric degree of freedom that allows them to minimize their energy when contrasting geometries are present in the problem ({\it e.g.}, electron band mass and dielectric tensor).
We investigate the internal metric of several incompressible FQH states by probing its response to band mass anisotropy using infinite DMRG simulations on a cylinder geometry. 
We test and apply a method to extract the internal metric of a FQH state from its guiding center structure factor.
We find that the response to band mass anisotropy is approximately the same for states in the same Jain sequence,
but changes substantially between different sequences. 
We provide a theoretical explanation of the observed behavior of primary states at filling $\nu = 1/m$ in terms of a minimal microscopic model of flux attachment.
\end{abstract}

\maketitle

\section{Introduction}

The internal geometry of fractional quantum Hall states\cite{Haldane2011} has attracted a great deal of interest recently, with the development of a variety of theoretical tools
such as anisotropic pseudopotentials\cite{HaldanePP, BoYang2017A, BoYang2017B} and model wavefunctions\cite{Qiu2012, Balram2016}, 
as well as field-theoretic techniques including gravitational anomalies\cite{Can2014, Can2015, Bradlyn2015}
and bi-metric geometry\cite{Gromov2017A, Gromov2017B}.
Areas that have seen important progress recently include the understanding of nematic and anisotropy-driven transitions\cite{You2014, Zhu2017, Zhu2018, Lee2018},
higher-spin modes of the composite Fermi liquid (CFL)\cite{Nguyen2017, Ippoliti2017C}, 
and out-of-equilibrium dynamics of the geometric degrees of freedom\cite{Liu2018}.

On the experimental side, the CFL state at filling $\nu = 1/2$ has been intensely studied.
Experiments in patterned GaAs quantum wells can detect commensurability oscillations in transport, which can be used to map the shape of the Fermi contour\cite{Gokmen2010, Kamburov2013}.
This has allowed the measurement of the CFL's response to different kinds of anisotropy, 
including strain-induced quadrupolar distortions\cite{Jo2017}, 
higher-moment warping of hole bands\cite{Mueed2015},
and splitting of hole bands into separated Fermi pockets\cite{Kamburov2014}.
In all three cases, measurements are found to agree with numerical simulations\cite{Ippoliti2017A, Ippoliti2017C, Ippoliti2017B}.

Incompressible states, lacking a Fermi contour, do not offer this experimental route to the internal geometry.
Transport measurements can provide information on symmetry-broken phases such as stripes\cite{Samkharadze2016}, but cannot directly provide information about anisotropic FQH states.
Though alternative ways to test the internal geometry of gapped states have been put forward, {\it e.g.} by using acoustic wave absorption\cite{Yang2016}, 
at this stage our knowledge comes almost exclusively from theory and numerics.

A considerable amount of work has been done on the geometry of gapped states, 
especially the state at filling fraction $\nu = 1/3$, 
where both numerical exact diagonalization and anisotropic model wavefunctions allow significant progress.
Problems that have been considered include 
the effect of anisotropic interaction\cite{Wang2012} or band mass\cite{BoYang2012}, 
in-plane magnetic fields\cite{Papic2013}, 
and a spatially-varying metric\cite{Johri2016}.

Being mostly limited to the $\nu = 1/3$ state, this body of work has left an important open question:
how does the response of a FQH state to anisotropy depend on the filling fraction $\nu$? 
What is the role of the underlying phase?
In this work, we address this question by performing extensive numerical simulations using the infinite density matrix renormalization group (iDMRG) method on incompressible states at various fillings, 
including multiple states in the first hierarchy/Jain sequence ($\nu = 1/3$, $2/5$ and $4/9$), and the state at $\nu = 1/5$.

The paper is organized as follows. 
In Sec.~\ref{sec:model} we introduce the model, its Hamiltonian and the relevant notation. 
In Sec.~\ref{sec:duncan} we present a detailed discussion of the guiding center structure factor, a key quantity for our numerical method, in the absence of rotational symmetry.
The numerical method itself is then explained in Sec.~\ref{sec:method}. 
Results are presented in Sec.~\ref{sec:numerical} and interpreted theoretically in terms of a minimal microscopic model of anisotropic flux attachment in Sec.~\ref{sec:theory}.
Finally, in Sec.~\ref{sec:discussion} we provide our conclusions and discussion.


\section{Model \label{sec:model}}

We consider the familiar setting for the fractional quantum Hall problem: 
a two-dimensional electron gas in a high perpendicular magnetic field $B$, in the presence of electron-electron interactions and in the absence of disorder.
The system is described by the following Hamiltonian:
\begin{equation}
H =  \sum_i  \frac{1}{2} \left( m^{-1} \right)^{ab} \pi_{i, a} \pi_{i, b} +
\sum_{i<j} V(|\mb r_i - \mb r_j|) \;,
\end{equation}
where $\bs \pi = \mb p - e \mb A$ is the dynamical momentum, $m^{-1}$ is the inverse mass tensor (generally anisotropic), and $V(r)$ is the isotropic Coulomb interaction.
Indices $i$ and $j$ enumerate the electrons in the system ($1, \dots N$), while $a$ and $b$ run over spatial components ($x$, $y$).
We can choose our coordinates so that the mass tensor is diagonal:
\begin{equation}
m_{ab} = m \begin{pmatrix} 1/\aE & 0 \\ 0 & \aE \end{pmatrix} \;,
\label{eq:alpha_0}
\end{equation}
where $\aE \equiv \sqrt{m_{yy}/m_{xx}}$ is a measure of the anisotropy\cite{Ippoliti2017A}.
The one-electron kinetic energy then becomes
\begin{equation}
H_0 = \frac{1}{2m} \left( \aE \pi_x^2 + \frac{1}{\aE} \pi_y^2 \right)  \;.
\end{equation}

We take the cyclotron gap $\hbar \omega_c$ to be much larger than all other energy scales in the problem, so that the dynamics can be projected into the lowest Landau level (LLL) and mixing with higher Landau levels can be safely neglected. 
The Hamiltonian projected in the LLL is
\begin{equation}
H_{LLL} = \sum_{i<j} \sum_{\mb q} \tilde{V}(\mb q) e^{i \mb q \cdot (\mb R_i - \mb R_j)}
\label{eq:hlll}
\end{equation}
where $\mb R_j$ is the guiding center operator for the $j$-th electron and
$$
\tilde{V}(\mb q) \equiv V(\mb q) |F_0(\mb q)|^2 
= \frac{2\pi}{q} e^{-\frac{1}{2} \ell_B^2 \left( q_x^2 \aE + q_y^2 /\aE \right)}
$$ is the \emph{effective interaction}, 
corresponding to the bare Coulomb interaction $V(\mb q)$ between two anisotropic LLL orbitals, whose shape is encoded in the form factor $F_0(\mb q)$.
Importantly, this effective interaction has two distinct metrics in it:
that of the dielectric tensor (which we choose to be $\varepsilon_{ab} \propto \delta_{ab}$) and that of the band mass $m_{ab}$. 
A coordinate rescaling can only move the anisotropy between these two metrics, but cannot remove it from the problem, which is irreducibly anisotropic (we will present a more rigorous definition of anisotropy in Sec.~\ref{sec:duncan}).

Previous investigations\cite{Ippoliti2017A} have focussed on the composite Fermi liquid state at filling $\nu = 1/2$, which has a Fermi contour that can be used to directly access the internal metric of the state.
Gapped fractions do not offer this possibility, so different methods have to be devised.
One possibility is to consider the overlap of numerically obtained ground states with anisotropic model wavefunctions\cite{BoYang2012}.
However, that is subject to the assumption that anisotropic model wavefunctions are a faithful description of the state, and becomes more complicated for states at $\nu \neq 1/m$.
In this work, we use a different approach which is enabled by our numerical method.
The \emph{infinite} cylinder geometry allows us to probe a \emph{continuum} of wavevectors along the cylinder axis.
This, in turn, gives us a new method to access the internal geometry of the state. 
The method, which we thoroughly describe in Section~\ref{sec:method}, is based on the calculation of the static guiding center structure factor.
In the absence of isotropy, this quantity contains important information about the geometry of the FQH state,
which we describe in the next Section.


\section{Guiding center structure factor of anisotropic FQH states}\label{sec:duncan}

The static guiding center structure factor can be defined as
\begin{equation}
S(\mb q) = \frac{1}{N_\phi} \langle \delta {\rho} (\mb q) \delta {\rho}(- \mb q) \rangle \;,
\label{eq:s_def}
\end{equation}
where $\rho(\mb q)$ is the Fourier transform of the guiding center density operator, 
\begin{equation}
{\rho} (\mb q) = \sum_i e^{i \mb q \cdot \mb R_i} \;.
\end{equation}
In Eq.~\eqref{eq:s_def}, we use the regularized operator
\begin{equation}
\delta {\rho} (\mb q) 
= {\rho}(\mb q) - \langle {\rho}(\mb q) \rangle
= {\rho}(\mb q) - 2\pi \nu \delta(\mb q \ell_B) \;,
\end{equation}
which represents fluctuations of the density away from the uniform background value of $\nu/2\pi\ell_B^2$.
A crucial property of $S(\mb q)$, related to the incompressibility of the state, is that its small-$q$ behavior is \emph{quartic}:
$S(u \mb q) \sim u^4$ as $u \to 0$. 
The long-wavelength structure of $S(\mb q)$ encodes important information about the geometry of the state\cite{Haldane2009}. 
In particular, for an incompressible, translationally invariant ground state, as $u \to 0$ one has
\begin{equation}
S(u \mb q) \sim \frac{1}{4} (u \ell_B)^4 \Gamma^{abcd} q_a q_b q_c q_d \;,
\label{eq:general_S}
\end{equation}
where the $\Gamma$ tensor is
\begin{align}
\Gamma^{abcd} 
& = \frac{1}{N_\phi} \left( \ev{\frac{1}{2} \{ \Lambda^{ab}, \Lambda^{cd} \} } 
- \ev{\Lambda^{ab}} \ev{\Lambda^{cd}} \right), 
\label{eq:gamma_def} \\
\Lambda^{ab} 
& = \frac{1}{2\ell_B^2} \sum_i \{ R^a_i, R^b_i \} \;.
\label{eq:lambda_def}
\end{align}
Equations~\eqref{eq:general_S}, \eqref{eq:gamma_def} and \eqref{eq:lambda_def} can be proven by starting from the following formula for the structure factor,
\begin{align}
S(\mb q) 
& = \frac{1}{N_\phi} \sum_{i, j} \langle e^{i\mb q \cdot \mb R_i} e^{-i\mb q \cdot \mb R_j} \rangle - \langle e^{i\mb q \cdot \mb R_i} \rangle \langle e^{-i\mb q \cdot \mb R_j} \rangle \;,
\end{align}
and expanding $S(u\mb q)$ in powers of $u$ around $u = 0$:
$$
S(u \mb q) = \sum_{k=0}^\infty u^k \sum_{m, n = 0}^\infty \delta_{m+n, k} C_{m, n} \;.
$$
The coefficients $C_{mn}$ are
\begin{align}
C_{mn} 
& \equiv \frac{i^{m}(-i)^n}{m!n! N_\phi }  \sum_{i, j} \langle  (\mb q \cdot \mb R_i)^m  (\mb q \cdot \mb R_j)^n \rangle  \nonumber \\
& \qquad - \langle  (\mb q \cdot \mb R_i)^m \rangle \langle (\mb q \cdot \mb R_j)^n \rangle \;.
\end{align}
All coefficients $C_{0,m}$ and $C_{m, 0}$ vanish trivially.
Coefficients $C_{1, m}$ and $C_{m, 1}$ also vanish, because $\sum_i R_i^a = -\ell_B^2 \epsilon^{ab} P_b$, where $P$ is the total momentum operator (generator of center-of-mass translations), 
which annihilates the translationally invariant ground state.
Therefore the lowest-order contributions is $O(u^4)$, coming from $m=n=2$:
\begin{align}
C_{2,2}  
& = \frac{\ell_B^4}{4N_\phi}q_a q_b q_c q_d
(\langle \Lambda^{ab} \Lambda^{cd} \rangle - \langle \Lambda^{ab} \rangle \langle \Lambda^{cd} \rangle)\;.
\end{align}
This finally yields
\begin{equation}
S(u\mb q) = \frac{(u\ell_B)^4}{4} \Gamma^{abcd} q_a q_b q_c q_d + O(u^6)
\end{equation}
with $\Gamma^{abcd}$ given by \eq{gamma_def}.

The symmetric matrix $\Lambda^{ab}$ consists of three independent Hermitian operators ($\Lambda^{xx}$, $\Lambda^{xy}$, $\Lambda^{yy}$) 
which generate area-preserving diffeomorphisms of the plane:
depending on the matrix of coefficients $\eta_{ab}$, the unitary
$$
U(\eta) \equiv e^{i \eta_{ab} \Lambda^{ab}} 
$$
can be a squeezing transformation ($\det \eta <0$), a shear transformation ($\det \eta = 0$) or a rotation ($\det \eta >0$).
The latter case allows for a general, coordinate-independent notion of rotational symmetry:
if there exists a metric $g$ such that
\begin{equation}
[\mc H, g_{ab} \Lambda^{ab} ] = 0\;,
\label{eq:general_isotropy}
\end{equation}
then the rotation group generated by the angular momentum operator
$$L^z(g) = g_{ab} \Lambda^{ab} $$ 
is a continuous symmetry of the system.
The system is thus isotropic
(though that might not be manifest in a coordinate system where $g_{ab} \neq \delta_{ab}$).
 
In the most general case, the tensor $\Gamma$ is defined by two unimodular metrics, $g_1$ and $g_2$, and two real coefficients $\kappa$, $\xi$ as follows:
\begin{equation}
\Gamma^{abcd} = \kappa(g_1^{ac} g_1^{bd} + g_1^{ad} g_1^{bc}) +\xi g_2^{ab} g_2^{cd} \;.
\label{eq:gamma_g}
\end{equation}
If there exists a metric $g$ that satisfies Eq.~\eqref{eq:general_isotropy}, then symmetry requires $g_1 = g_2 = g$.
Furthermore, since the ground state is an eigenstate of $L^z(g)$, one has
$$
g_{ab} \Gamma^{abcd} = \frac{1}{N_\phi} \left( \ev{\{L^z(g), \Lambda^{cd} \}/2} - \ev{L^z(g)} \ev{\Lambda^{cd}}\right)=0
$$
which forces $\xi = -\kappa$, giving
\begin{equation}
\Gamma^{abcd} = \kappa(g^{ac} g^{bd} + g^{ad} g^{bc} - g^{ab} g^{cd}) \;.
\end{equation}
Under this assumption of isotropy, the quartic part of the structure factor becomes the perfect square of a quadratic:
$$
S(u \mb q) \sim \frac{\kappa}{4} (u \ell_B)^4 \left(|\mb q|^2_g\right)^2\;,
$$
where $|\mb q|_g^2 \equiv g^{ab} q_a q_b$.
In general, the metrics $g_1$ and $g_2$ need not be the same, 
and the general form of the long-wevelength structure factor is an arbitrary quartic. 
Additional symmetries (e.g. reflection or discrete rotations) can put restrictions on it.


\section{Numerical method and symmetry considerations \label{sec:method}}

We use an infinite density matrix renormalization group (iDMRG) algorithm for quantum Hall states\cite{Zaletel2013, Zaletel2015}. 
The algorithm uses an infinite cylinder geometry, which allows a mapping to a one-dimensional fermion chain. 
Varying the cylinder circumference $L$ allows control over finite-size effects.
We pick a coordinate system with the $\hat x$ axis along the infinite direction of the cylinder 
and the $\hat y$ axis along the circumference, 
so the $q_y$ component of momentum can take discrete values $2\pi n/L$, $n \in \mathbb Z$, while $q_x$ is in principle continuous.

\begin{figure}
\centering
\includegraphics[width = 0.95\columnwidth]{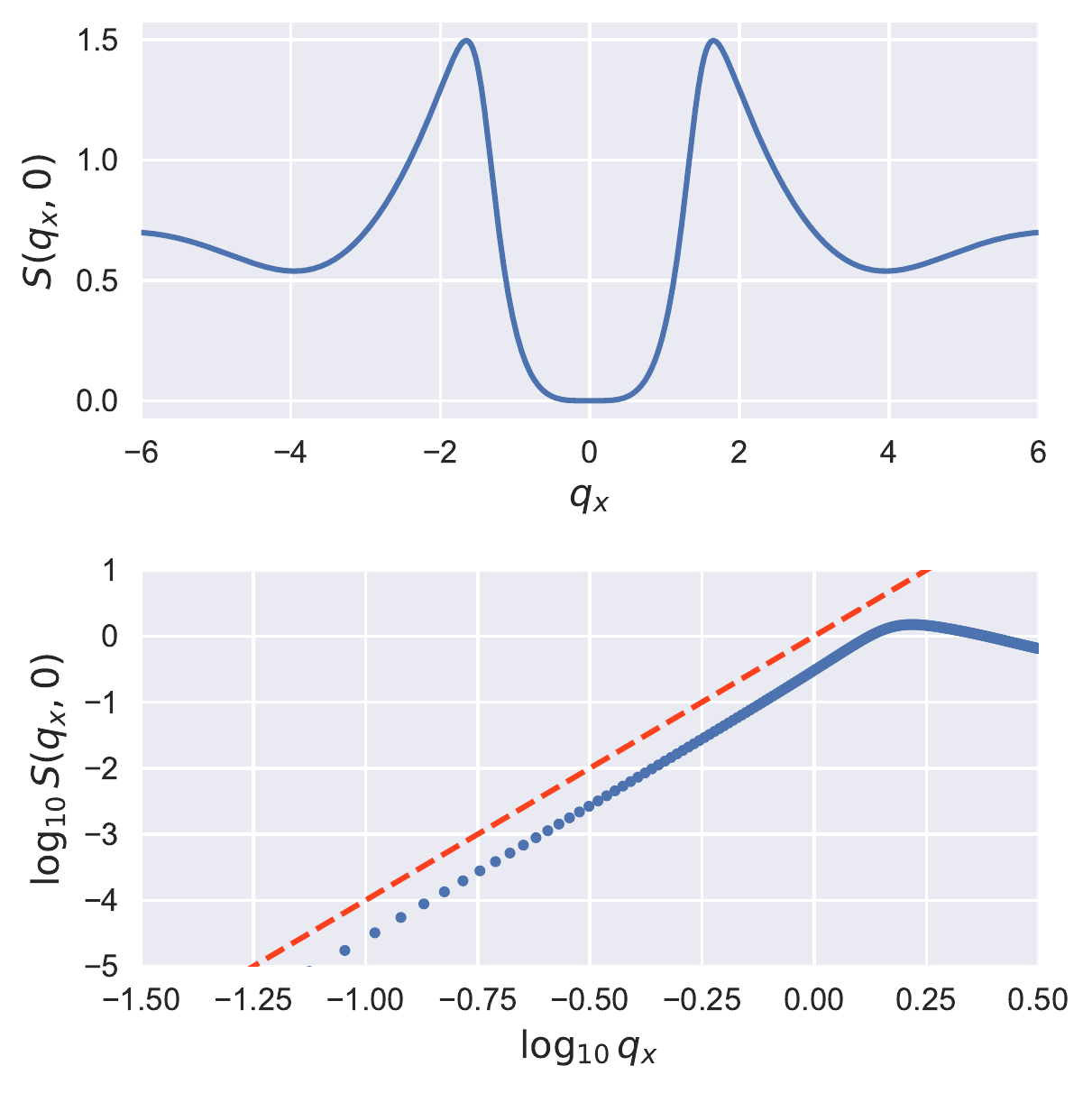}
\caption{Top: structure factor $S(q_x, 0)$ for the $\nu = 1/3$ state, computed numerically with iDMRG on a cylinder with circumference $L = 18\ell_B$, at bond dimension $\chi = 1024$.
Bottom: same quantity on a logarithmic scale (dots) compared to the quartic monomial $q_x^4$ (dashed line, slope 4).}
\label{fig:S_onethird}
\end{figure}

We parametrize the electron mass tensor as
\begin{equation}
m_{ab} = m \begin{pmatrix} e^{-\x} & 0 \\ 0 & e^{\x} \end{pmatrix}\;,
\qquad \x \in \mathbb R\;.
\end{equation}
This is parametrization is related to \eq{alpha_0} by $\x = \ln \aE$,
and has the advantage that a $\pi/2$ rotation acts simply as $\x \mapsto -\x$ (there are infinitely many such parametrizations; we choose this one for convenience).

At any of the fillings discussed below, we compute the matrix product state (MPS) approximation to the many-body ground state for a range of values of the cylinder circumference $L$ and the anisotropy parameter $\x$.
For each ground state, we compute the guiding center structure factor $S(q_x, 0)$. 
Details about the calculation of $S(\mb q)$ from the MPS can be found in Ref.~\cite{Geraedts2016}. 
An example for the $\nu = 1/3$ state is shown in Fig.~\ref{fig:S_onethird}, where the quartic behavior $S(q_x, 0) \sim q_x^4$ is clearly visible near the origin.

\begin{figure}
\centering
\includegraphics[width=\columnwidth]{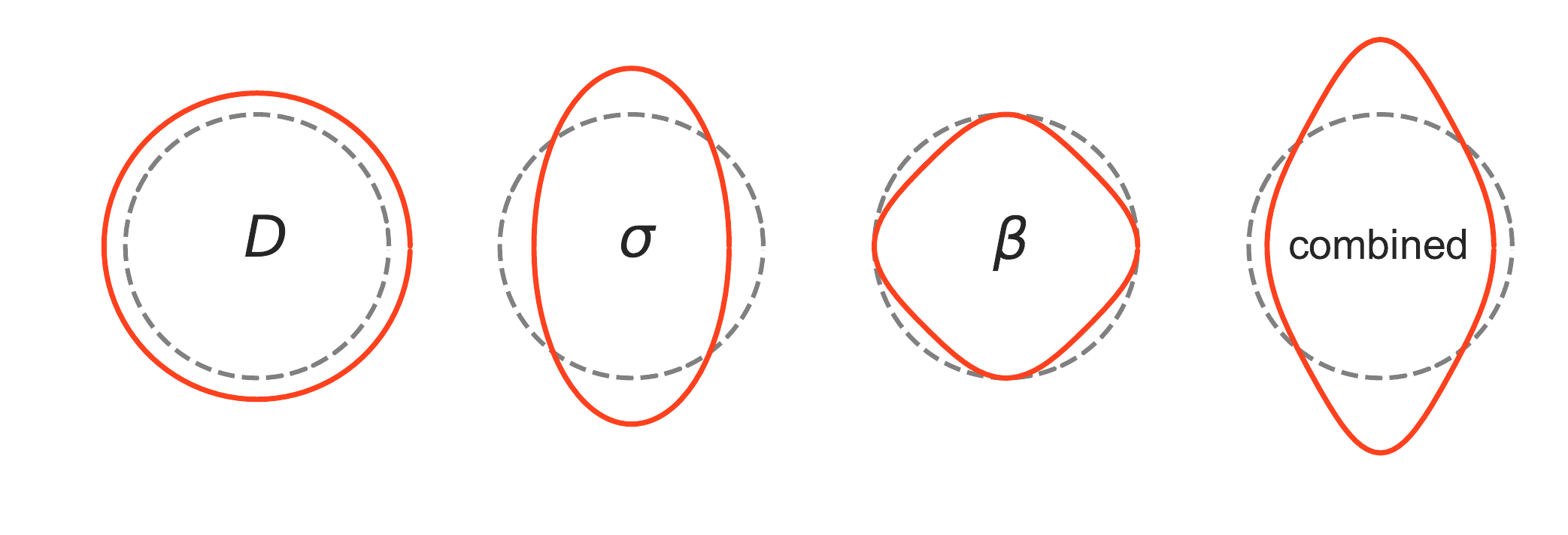}
\caption{Effect of the terms in \eq{s_reparam} on the shape of equal-value contours of the guiding center structure factor $S(q)$ near $q=0$.
From left to right: $\D$ parametrizes an isotropic rescaling;
$\y$ a unimodular metric (uniaxial stretching);
$\beta$ a $C_4$-symmetric distortion.
An example where all three are present is shown on the right.
The dashed circle corresponds to $\D = \y = \beta = 0$, for comparison.
\label{fig:shapes}
}
\end{figure}

Based on the discussion in the previous Section, 
and taking into account the reflection symmetries $q_x \leftrightarrow -q_x$, $q_y \leftrightarrow -q_y$, 
the long-wavelength limit of the structure factor must take the quartic form
\begin{equation}
S(u \mb q) \sim u^4(A q_x^4 + B q_x^2 q_y^2 + C q_y^4)
\end{equation}
for $u \to 0$.
A convenient re-parametrization of the coefficients $A$, $B$, $C$ is as follows:
$$
A \equiv e^{2(\D+\y)},
\qquad
B \equiv 2e^{2\D}(1+\beta),
\qquad
C \equiv e^{2(\D-\y)}\;,
$$
which gives
\begin{equation}
S(u\mb q) \sim u^4 e^{2\D} \left( (e^\y q_x^2 + e^{-\y} q_y^2)^2 + 2\beta q_x^2 q_y^2 \right)\;.
\label{eq:s_reparam}
\end{equation}
A $\pi/2$ rotation acts by mapping $\x\mapsto - \x$ while exchanging $q_x^2$ and $q_y^2$, 
therefore the parameters transform as follows:
\begin{equation}
\begin{aligned}
\D(-\x) & = \D(\x), \\
\y(-\x) & = -\y(\x), \\
\beta(-\x) & = \beta(\x)\;.
\end{aligned}
\end{equation}
The physical meaning of these parameters is apparent from \eq{s_reparam}:
$\D$ describes a change of the overall magnitude of $S(\mb q)$, in an isotropic fashion; 
$\y$ parametrizes an internal unimodular metric of the FQH state that describes a uniaxial squeezing;
and $\beta$ represents a possible $C_4$-symmetric anisotropy that remains once a coordinate transformation $q_x \mapsto e^{-\y/2} q_x$, $q_y \mapsto e^{\y/2} q_y$ removes the uniaxial squeezing.
This is illustrated graphically in Fig.~\ref{fig:shapes}.

If one assumes the absence of the $C_4$-symmetric distortions parametrized by $\beta$, the long-wavelength limit of the structure factor assumes the particularly simple form of a perfect square,
\begin{equation}
S(u\mb q) \sim u^4 (g^{ab} q_a q_b)^2, 
\quad
g^{ab} = \text{diag}(e^{\D+\y}, e^{\D-\y})\;.
\label{eq:perfect_square}
\end{equation}
In Appendix~\ref{app:check} we perform a numerical test of this hypothesis on the $\nu = 1/3$ state by tilting the band mass tensor relative to the cylinder axis, 
effectively allowing us to probe the $q_x = \pm q_y$ directions in \eq{s_reparam},
and thus gain information on the $\beta$ coefficient.
Our test indicates that $\beta = 0$ within our numerical and finite-size accuracy. 
Therefore, all the information about the long-wavelength limit of $S(\mb q)$ is contained in the parameters $\D$ and $\y$.
These can be extracted from a quartic fit of $S (q_x, 0)$ at small $q_x$, while keeping the band mass tensor aligned with the cylinder.
We have
\begin{equation}
S(q_x, 0) \simeq \lambda(\x) q_x^4 \equiv e^{2(\D(\x)+\y(\x))} q_x^4\;,
\label{eq:fit_lambda}
\end{equation}
and by exploiting the different parity of $\D(\x)$ and $\y(\x)$ we obtain
\begin{align}
\y(\x) = \frac{1}{4} \log \frac{\lambda(\x)}{\lambda(-\x)}\;,
\quad
\D(\x) = \frac{1}{4} \log \lambda(\x) \lambda(-\x)\;.
\label{eq:naive}
\end{align}
On a cylinder with finite circumference, the values of $\x$ and $-\x$ need not be related by an exact duality, as the boundary conditions explicitly break $C_4$ rotation symmetry.
In that case, \eq{perfect_square} does not rigorously hold and there can be more information in the geometry of the state which is not captured by \eq{naive}.
Therefore, care must be taken to analyze finite-size effects appropriately.

\begin{figure}
\centering
\includegraphics[width = 0.95\columnwidth]{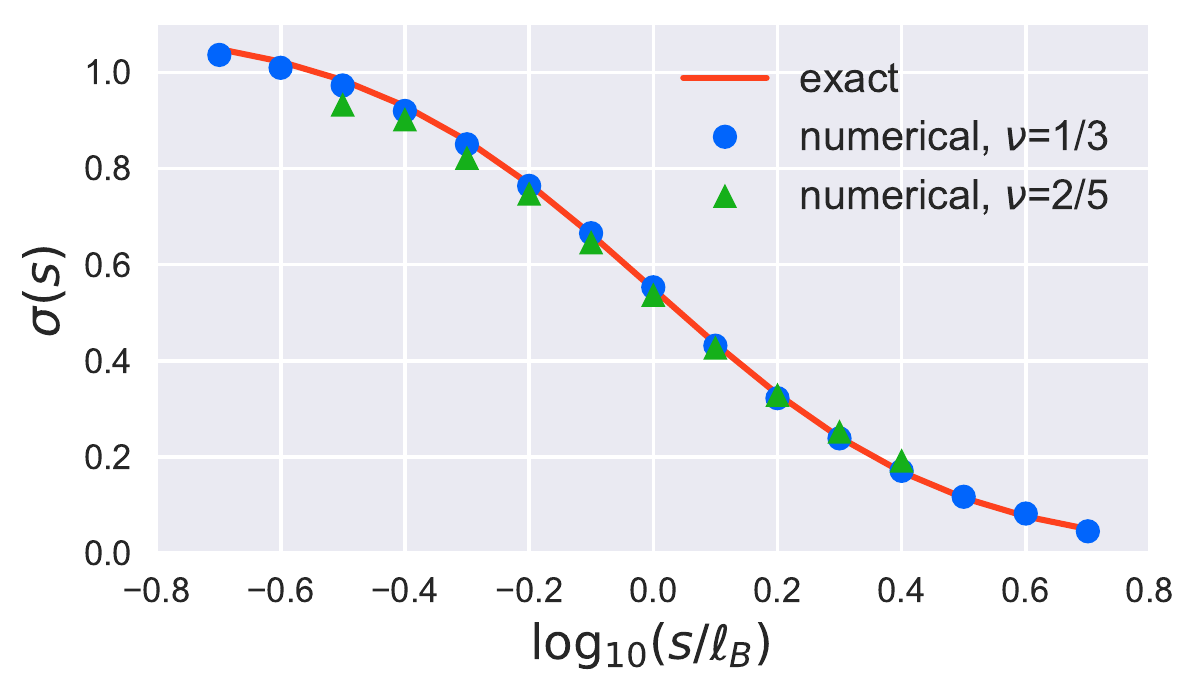}
\caption{
Comparison between numerical method and exact result for Gaussian electron-electron interaction\cite{Yang2013}. 
The numerical data is obtained with iDMRG on a cylinder with circumference $L = 18\ell_B$ for filling fractions $\nu = 1/3$ and $2/5$. The bond dimension is $\chi = 1024$.
We find the ground states for $\x = \log 3$ and use Eq.~\eqref{eq:naive} to estimate $\y$.
For the $\nu = 1/3$ state the agreement is nearly perfect in the entire range of $s$, whereas for the $\nu = 2/5$ state there are small but visible deviations at the largest values of $\y$ (small $s$). 
\label{fig:ky}
}
\end{figure}

In Fig.~\ref{fig:ky} we benchmark this method against the known exact result\cite{Yang2013} for Gaussian electron-electron interaction 
$V(r) = e^{-\frac{1}{2}(r/s)^2}$ with interaction length scale $s$.
This problem is actually isotropic:
it is invariant under the generator $L^z(g) = g_{ab} \Lambda^{ab}$ for the metric 
$g_{ab} = \text{diag}(e^\y, e^{-\y})$, with 
$$
\y = \frac{1}{4} \log \frac{e^{2\x} + s^2/\ell_B^2}{e^{-2\x} + s^2/\ell_B^2} \;.
$$
We numerically compute the interacting ground states at $\nu = 1/3$ and $\nu = 2/5$ in a cylinder with circumference $L = 18 \ell_B$ for electron mass anisotropy parameter $e^\x = 3$ and a wide range of interaction length scales $s$.
We find very good agreement between our estimate of $\y$ and the exact result, showing how in this simple case the $L \to \infty$ formulas appear to be valid for all practical purposes.

Throughout the rest of the paper, 
we study genuinely anisotropic FQH states that in some cases exhibit significant finite-circumference effects.
In order to take into account the effects of finite $L$, it is helpful to 
assume smoothness around the isotropic point $\x = 0$, and Taylor expand
\begin{equation}
\log(\sqrt{\lambda}) = \D + \y =  \sum_n c_n \x^n \;.
\label{eq:taylor}
\end{equation}
By construction, the even-$n$ terms contribute to $\D$ and the odd-$n$ ones contribute to $\y$.
By fitting $\log \sqrt{\lambda}$ to a polynomial at each size, it is possible to analyze the finite-size drift of the contributions to $\D$ and $\y$ and infer their $L \to \infty$ limit.
In all cases we present in the following, 
keeping only terms with $n\leq 3$ is sufficient to give a good fit to the data.
Further, we find $c_2$ to become consistent with 0 as $L$ is increased, 
indicating that the overall scale parameter $\D$ is compatible with a constant.
Coefficients $c_1$ and $c_3$ characterize the dependence of the internal metric parameter $\y$ on the electron anisotropy $\x$ and are found to approach finite values as $L \to \infty$.


\section{Numerical results \label{sec:numerical}}

\subsection{First Jain sequence: $\nu = 1/3$, $2/5$, $4/9$\label{sec:13}}

\begin{figure}
\centering
\includegraphics[width = 0.99\columnwidth]{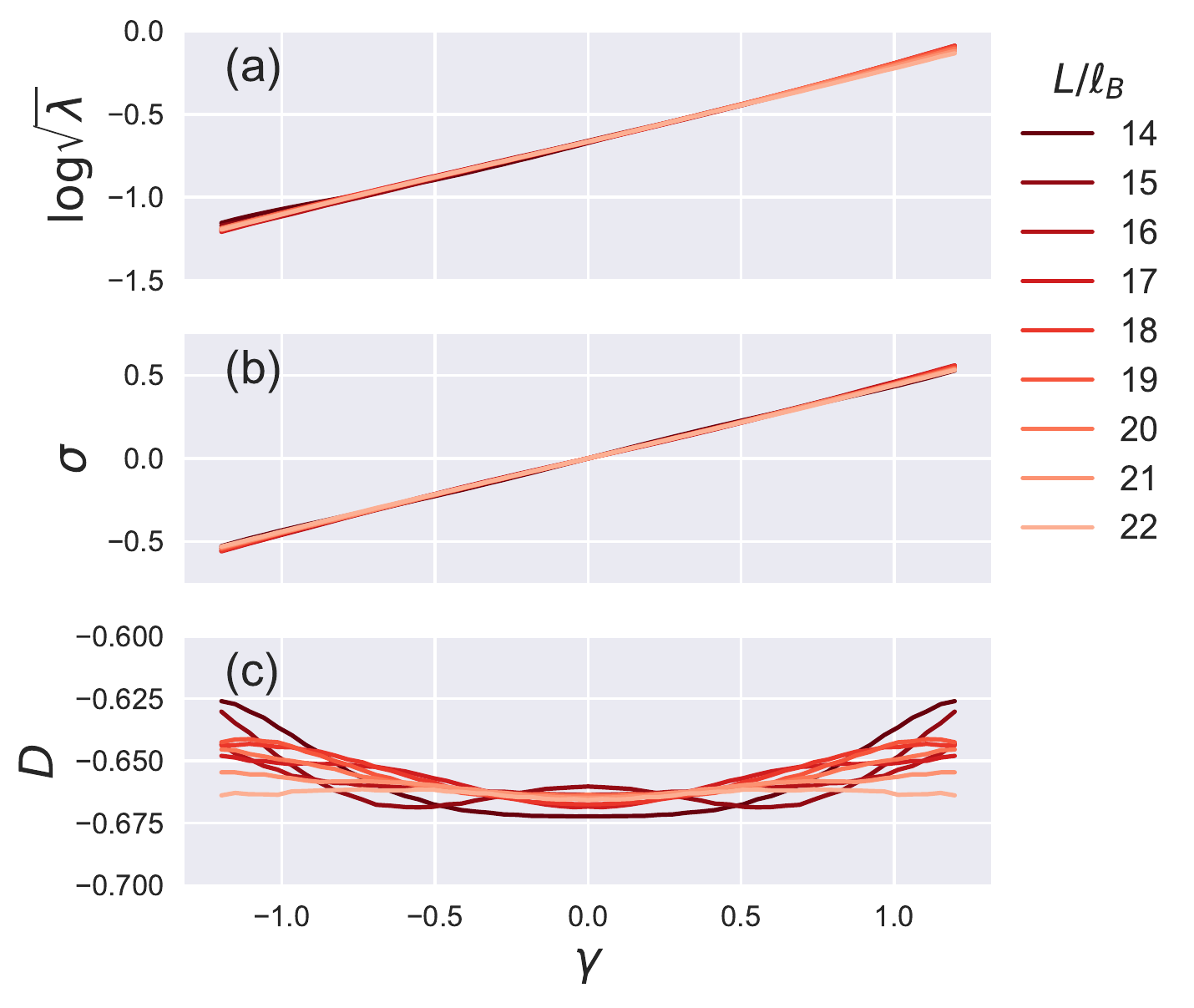} 
\caption{
Numerical data for the $\nu = 1/3$ state as a function of electron anisotropy $\x$ and cylinder circumference $L$.
The data is obtained with iDMRG at bond dimension $\chi = 1024$.
(a) Quartic coefficient $\lambda(\x)$, extracted from $S(q_x, 0) \approx \lambda(\x) q_x^4$.
(b) $\y(\x)$, defined as the odd part of $\log\sqrt{\lambda}$.
(c) $\D(\x)$, defined as the even part of $\log\sqrt{\lambda}$, has small finite-size fluctuations and approaches a constant as $L$ increases.
\label{fig:13coul}
}
\end{figure}

We start from the FQH state with the largest gap, the one at filling $\nu = 1/3$.
We gather data for 50 equally spaced values of $\x$ in the range $-1.2 \lesssim \x \lesssim 1.2$ (corresponding to $0.3\lesssim \aE \lesssim 3.3$) and cylinder circumferences $14\ell_B \leq L \leq 22\ell_B$.
In Fig.~\ref{fig:13coul} we plot the value of $\log \sqrt{\lambda(\x)}$, 
which is equal to $\y + \D$ as discussed in the previous section,
as a function of $\x$ for all the circumference sizes.
As can be seen, finite-size effects are rather small and are stronger for $\x<0$, 
where correlations are elongated in the direction of the finite circumference.

\begin{figure}
\centering
\includegraphics[width = 0.45\textwidth]{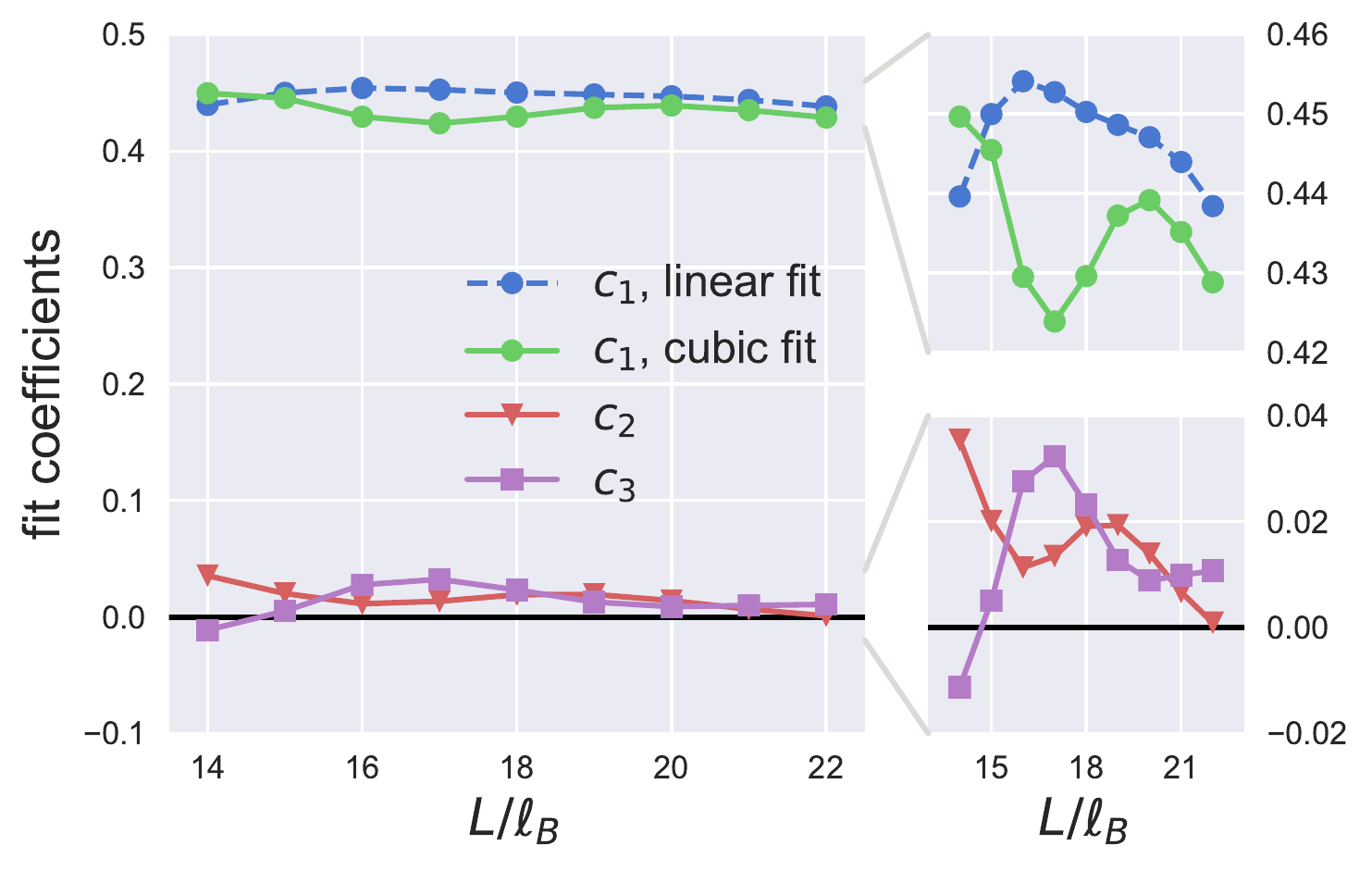} 
\caption{
Coefficients for polynomial fit of $\log(\sqrt{\lambda(\x)})$ as a function of $\x$ for the $\nu = 1/3$ (data in Fig.~\ref{fig:13coul}).
Left: all coefficients from linear and cubic fits shown together. 
$c_1$ is more than an order of magnitude larger than all higher-order terms.
Right: $c_1$ and $c_{2,3}$ coefficients shown separately for clarity. 
$c_2$ is found to oscillate around $c=0$ (black line), consistent with Fig.~\ref{fig:13coul}.
\label{fig:coeffs13}
}
\end{figure}

Following the scheme described in the previous Section, and in particular \eq{taylor},
we fit $\log \sqrt{\lambda(\x)}$ to a cubic polynomial in $\x$:
\begin{equation}
\log \sqrt{\lambda(\x)}= c_0 + c_1 \x + c_2 \x^2 + c_3 \x^3 \;.
\label{eq:polyfit}
\end{equation}
In the planar limit ($L \to \infty$), coefficients $c_0$ and $c_2$ represent contributions to \D(\x), 
while $c_1$ and $c_3$ contribute to \y(\x).
Though this is only expected to hold at infinite $L$, 
by probing a range of circumference sizes and performing the fit procedure in \eq{polyfit} at each size we can infer the $L \to \infty$ limit.
We show the results in Fig.~\ref{fig:coeffs13}. 
As can be seen, $c_2 \to 0$ as $L$ increases, while $c_1$ and $c_3$ appear to go to non-zero values,
with $c_3$ being very small.
Thus, in the planar limit $L \to \infty$, we get
\begin{equation}
\y \simeq 0.43 \x + 0.01 \x^3 \;,
\quad \D \simeq \text{const.}
\end{equation}
(notice the absence of the $\x^2$ term in \D).
Even for the smallest sizes, $\D$ is constant within $\sim 4\%$, while $\y$ is approximately linear in $\x$.
We also show the results of a linear fit, $\y = c_1 \x$ with $c_3$ set to zero.
This would correspond to a power-law relation between the band mass metric and the internal FQH state metric, $g_{\sf{FQH}} \propto (g_m)^{c_1}$.
The result of this fit is $c_1 \simeq 0.445$.

We repeat the same analysis on the states at filling $\nu = 2/5$ and $\nu = 4/9$.
These are ``daughter states'' of the $\nu = 1/3$ state in the hierarchy scheme\cite{Haldane1983}, 
while in the composite fermion picture\cite{Jain1989, JainBook} they correspond to different numbers of filled ``$\Lambda$ levels'' in the first Jain sequence.
We obtain similar numerical results, which are shown in detail in Appendix~\ref{app:more_numerics}.
These states have smaller energy gaps, which imply longer correlation lengths, resulting in stronger finite-size effects.
Thus, especially for the $\nu = 4/9$ state, the oscillations of the $c_2$ and $c_3$ coefficients as a function of size still have a significant amplitude at the largest size we consider, $L = 22\ell_B$.
This increases the uncertainty on the estimate of $c_1$, as well.
Nonetheless, the results for both states appear compatible with the $\nu = 1/3$ ones.

\subsection{Second Jain sequence: $\nu = 1/5$}

Having found the same result for all the states derived from $\nu = 1/3$, 
a natural question to ask is whether the response to applied anisotropy changes for states that are \emph{not} derived from $\nu = 1/3$. We consider the simplest such state, $\nu = 1/5$.
In terms of composite fermions, this belongs to a different Jain sequence, where flux attachment combines each electron to four flux quanta, rather than two.
In the hierarchy picture, this is the parent state of a different tree of daughter states.

\begin{figure}
\centering
\includegraphics[width = 0.99\columnwidth]{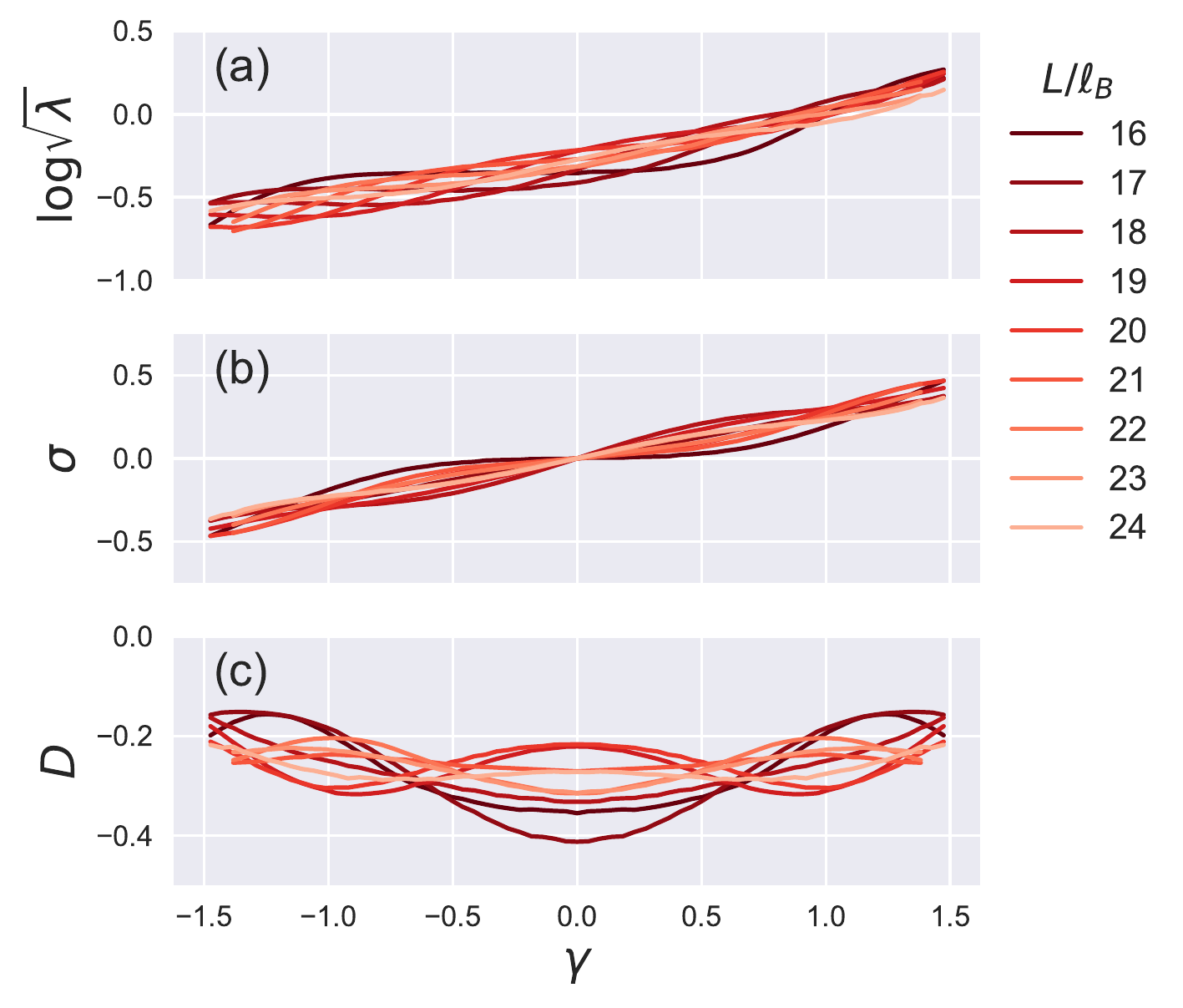}
\caption{
Same plots as Fig.~\ref{fig:13coul}, but for the $\nu = 1/5$ state.
Despite stronger finite-size effects, this state exhibits similar qualitative behavior as the $\nu  =1/3$ state.
\label{fig:15coul}
}
\end{figure}

In Fig.~\ref{fig:15coul} we show data collected for the state at $\nu = 1/5$ for cylinder circumferences $L = 16 \ell_B$ to $24\ell_B$.
Here, finite-size effects are more severe than for the $\nu = 1/3$ state,
requiring somewhat larger values of the cylinder circumference $L$ and a wider range of electron anisotropies $\x$.

\begin{figure}
\centering
\includegraphics[width = 0.9\columnwidth]{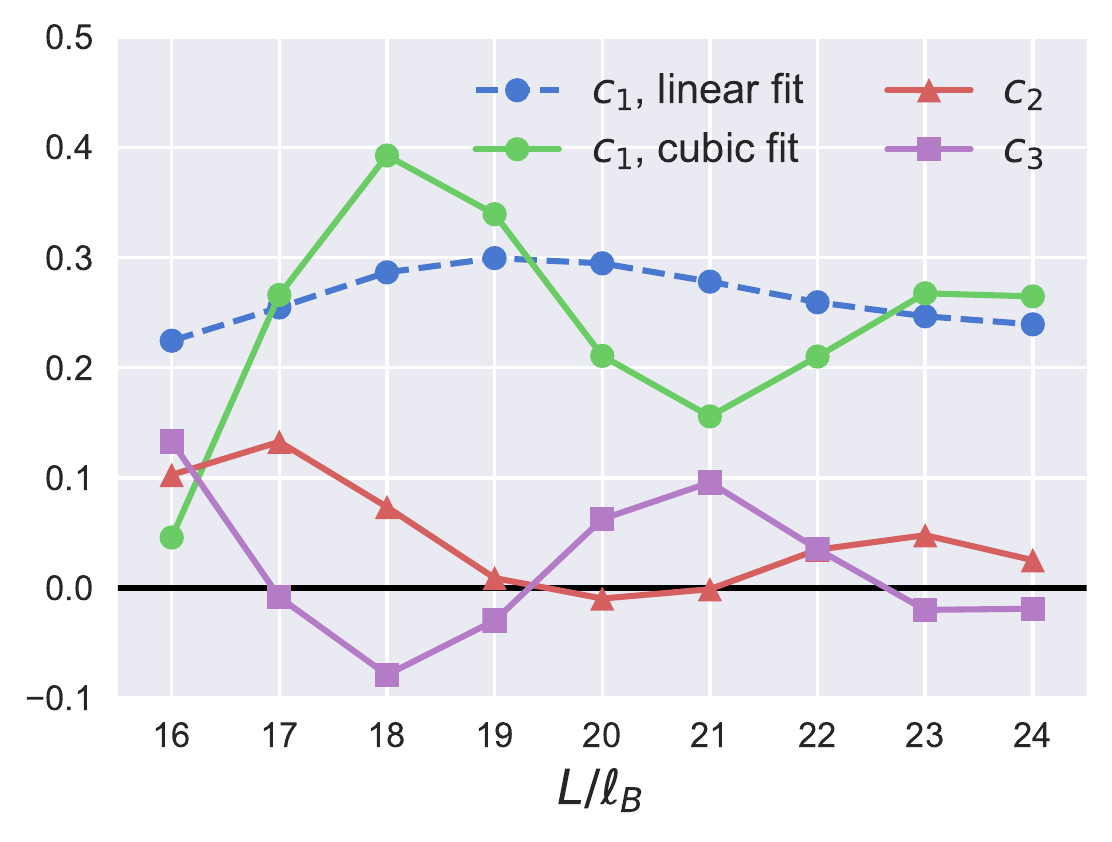}
\caption{
Coefficients for polynomial fit of $\log(\sqrt{\lambda(\x)})$ as a function of $\x$ for the $\nu = 1/5$ state (data in Fig.~\ref{fig:15coul}). The black line highlights $c=0$.
The behavior is qualitatively the same as that of the states in the first Jain sequence, albeit with fluctuations persisting to larger sizes.
Quantitatively, both estimates of $c_1$ are consistently below 0.4, and appear to stabilize between $0.25$ and $0.3$, a much smaller value than the $\nu = 1/3$ state.
\label{fig:coeffs15}
}
\end{figure}

Again, we define $\D + \y = \log \sqrt{\lambda}$ and perform a polynomial fit of this quantity as a function of $\x$.
The fit coefficients, shown in Fig.~\ref{fig:coeffs15}, exhibit wide fluctuations with system size, 
which have not yet decayed at circumference $L = 24\ell_B$. 
Nonetheless, the value of the $c_1$ coefficients, obtained from either a cubic or a linear fit of the data in Fig.~\ref{fig:15coul}, 
is consistently below 0.4 and appears to approach an asymptotic value between $0.25$ and $0.30$. 
We can therefore conclude that,
for a fixed value of the electron mass anisotropy $\x$,
the $\nu = 1/5$ state is much less anisotropic than states in the first Jain/hierarchy sequence.

\section{Microscopic model of flux attachment \label{sec:theory}}

In order to understand the observed response of states at different filling to applied band mass anisotropy, we consider a minimal microscopic model of flux attachment.
We consider two electrons with anisotropic band mass $m_{ab}$ in the lowest Landau level interacting with each other via the Coulomb interaction $V(r) = 1/r$. 
The system is  governed by an effective interaction $\tilde{V}(\mb r)$ which is the Fourier transform of
\begin{equation}
\tilde{V}(\mb q) 
= \frac{2\pi}{q} e^{-\frac{1}{2}\ell_B^2 (e^\x q_x^2 + e^{-\x} q_y^2) } \;.
\end{equation}
The two-body problem can be reduced to a single-particle problem in the relative coordinate, which describes an electron orbiting around a fixed potential $\tilde{V}(\mb r)$ generated by an anisotropic cloud of charge pinned to the origin.
The $m$ highest-energy orbitals of this potential, {\it i.e.} the bound states of $-\tilde{V}(\mb r)$, constitute an approximation of the ``excluded region'' around each electron in the incompressible many-body state at filling $\nu = 1/m$.
This is in analogy with the isotropic Laughlin wavefunction, where each factor of $(z_i-z_j)^m$ can be interpreted as arising from the exclusion of orbitals $z^0, \cdots z^{m-1}$ in the relative coordinate problem.

This simple model makes some unambiguous predictions. The effective interaction $\tilde{V}$ is isotropic at large distance (small $q$) and becomes more anisotropic at shorter distance (large $q$).
Therefore, the innermost orbitals in the excluded region are the most anisotropic, and as one moves outwards the orbitals become more and more isotropic.
Overall, the anisotropy of the composite boson is expected to \emph{decrease} with $m$.
This is explored in more detail in Appendix~\ref{app:veff}, and is in qualitative agreement with our numerical results from Section~\ref{sec:numerical}.

To be more quantitative, we 
start from the LLL Hamiltonian in Eq.~\eqref{eq:hlll} for two electrons,
\begin{equation}
\mathcal H_{\rm 2-body} = \sum_{\mb q} \tilde V(\mb q) e^{i \mb q \cdot (\mb R_1-\mb R_2) }  \;.
\end{equation}
We define the center-of-mass and relative guiding center coordinates as
\begin{equation}
R_{CM} \equiv \frac{R_1 + R_2}{\sqrt 2}\;,
\qquad 
\delta R \equiv \frac{R_1 - R_2}{\sqrt 2}\;,
\end{equation}
respectively.
Notice the normalization, which is necessary to maintain the correct canonical algebra, {\it i.e.} $[\delta R^x, \delta R^y] = -i\ell_B^2$.
The center of mass coordinate has no dynamics; we are thus left with a single-particle Hamiltonian for the relative motion:
$$
\mathcal H_{\text{rel.}}
= \sum_{\mb q} \tilde V(\mb q)  e^{i \mb q \cdot \delta \mb R \sqrt{2}}  
$$
Redefining the summation variable $\mb q \mapsto \sqrt{2} \mb q$, we get
\begin{equation}
\mathcal H_{\text{rel.}}
= \sum_{\mb q} \tilde{V}(\mb q/\sqrt{2})  e^{i \mb q \cdot \delta \mb R}  
\equiv \tilde{V}(\delta \mb R)\;.
\label{eq:relative}
\end{equation}
The factor of $\sqrt{2}$ is important because $\tilde{V}$ is not a homogeneous function.
In particular, it is expected to reduce anisotropy, as $\lambda \tilde{V}(\mb q/\lambda)$ approaches the isotropic Coulomb interaction as $\lambda \to \infty$.

A first approximation of the geometry of the excluded region can then be obtained by looking at equipotential contours of $\tilde{V}(\delta \mb R)$. 
Contours that enclose an area of $2\pi \ell_B^2(n+1/2)$ should indeed approximate the semiclassical electron trajectories.
We follow this approach in Appendix~\ref{app:veff}, finding it to give the correct qualitative dependence on $m$ and $\alpha$, but the quantitative agreement is not satisfactory.
This is not surprising, as the semiclassical approach works best for large $n$, while we are interested in the first few orbitals.
To get a more accurate quantitative description, one has to find the shape of the actual quantum eigenstates of $\tilde{V}$.

\begin{figure}
\centering
\includegraphics[width = 0.99\columnwidth]{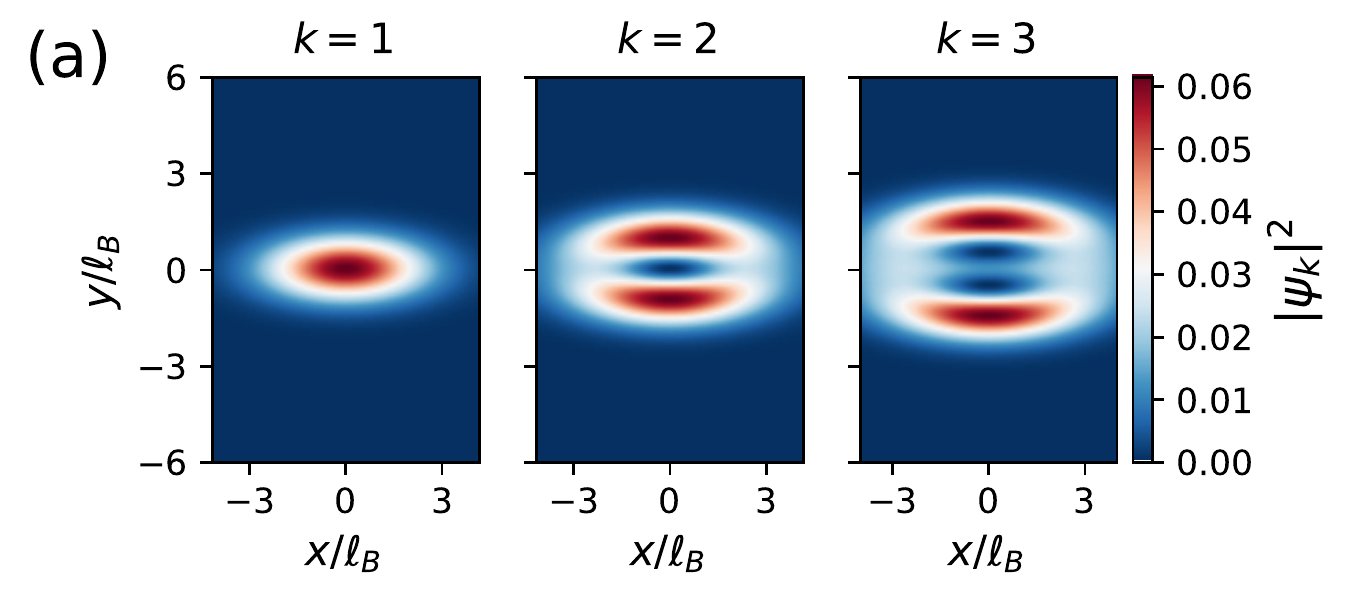} \\
\includegraphics[width = 0.99\columnwidth]{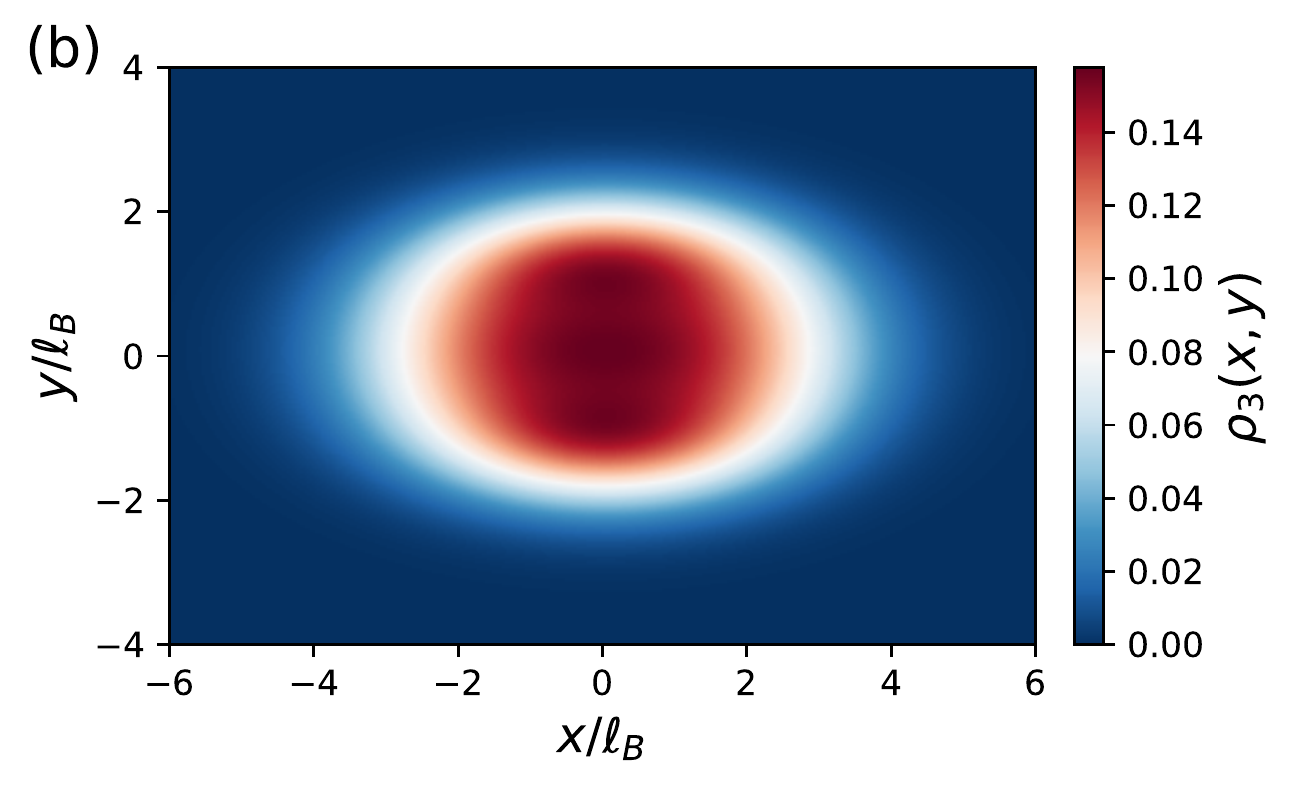} \\
\includegraphics[width = 0.99\columnwidth]{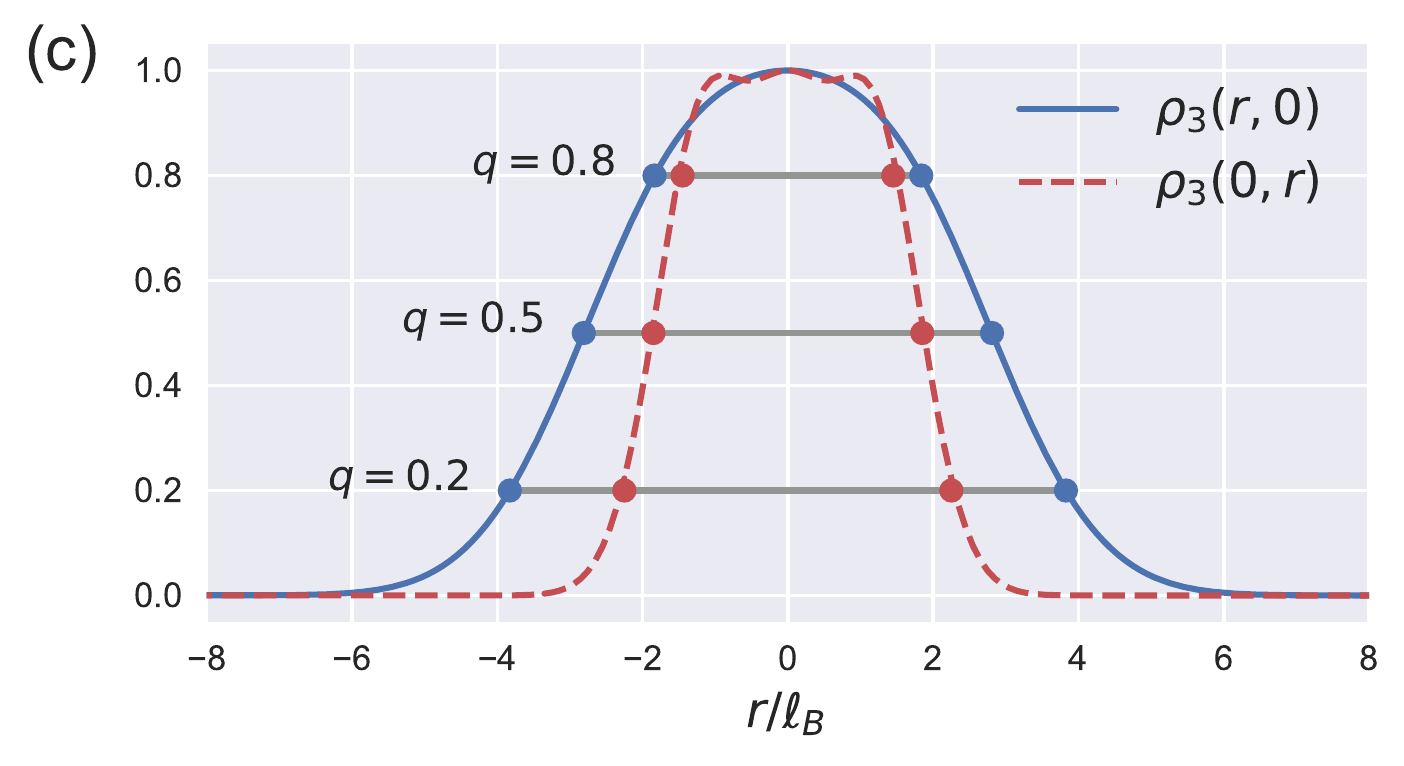}
\caption{
(a) Probability density of the three highest-energy orbitals $\psi_k$ of the relative Hamiltonian in \eq{relative} for $\x = \log 3$.
The orbitals have been calculated numerically on a torus with side $L = \sqrt{512\pi} \ell_B \simeq 40 \ell_B$, where finite-size effects are completely negligible.
(b) Probability density $\rho_m$ defined in Eq.~\eqref{eq:rhom}, for $m=3$.
The data are obtained by summing the three orbital probability densities above.
(c) Profiles of $\rho_3(x,y)$ along the lines $x=0$ and $y=0$, normalized so the maximum height is 1.
The two shapes are clearly different and no single rescaling can collapse them.
In particular, the ratio of widths at a fraction $q$ of the maximum height,
defined as $\estimate{q}$ in Eq.~\eqref{eq:estimate}, depends significantly on $q$ (we highlight $q=0.2$, $0.5$ and $0.8$ as examples).
\label{fig:prob}}
\end{figure}

We study this single-particle problem numerically on a torus with sides large enough that finite-size effects are negligible 
(a torus with side $L \simeq 40\ell_B$ is enough for this purpose) and numerically obtain eigenvalues $\{E_k\}$ and eigenvectors $\{ |\psi_k \rangle \}$\;.
We then calculate the real-space total probability density of the $m$ highest-energy wavefunctions,
\begin{equation}
\rho_m (x, y) = \frac{1}{m} \sum_{k = 1}^m |\psi_k(x, y)|^2 \;.
\label{eq:rhom}
\end{equation}
The result for $\alpha = 3$ is shown in Fig.~\ref{fig:prob} (a,b).

If $\tilde{V}(\mb q)$ is obtained by stretching a function of $q^2$, 
implying the problem is actually isotropic with respect to some metric $g_{ab} \neq \delta_{ab}$ as in the case of the Gaussian electron-electron interaction used in Fig.~\ref{fig:ky}, 
then $\rho_m(x,y)$ is also isotropic with respect to the same $g_{ab}$.
Therefore its profiles along the $y=0$ and $x=0$ lines, $\rho_m(r, 0)$ and $\rho_m(0,r)$, should have the same shape up to a rescaling factor $a$:
$$
\rho_{m} ( r/a, 0 )  = \rho_{m} (0, r a ) \quad \forall\ r\in \mathbb R \;.
$$
In that case the parameter $\y$ is unambiguously determined by $e^\y = \sqrt{a}$. 
For Coulomb interaction, though, the problem is genuinely anisotropic.
We find that the two profiles have significantly different shapes, as demonstrated in Fig.~\ref{fig:prob} (c), and no such rescaling exists.
In particular, one can estimate $\y$ as the ratio of the widths of the probability density profiles at various fractions $q$ of their height:
\begin{equation}
\estimate{q} \equiv \frac{y_q}{x_q}, \quad \rho_{m}(x_q, 0) = \rho_{m}(0, y_q) = q\rho_m(0,0)\;.
\label{eq:estimate}
\end{equation}
Each value $0<q<1$ defines an estimate $\estimate{q}$ of $\y$.
But the estimates so defined are found to depend significantly on $q$.
This might seem to make the model not predictive.
The situation turns out to be much better than this pessimistic outlook, as we now discuss.

\begin{figure}
\centering
\includegraphics[width = 0.95 \columnwidth]{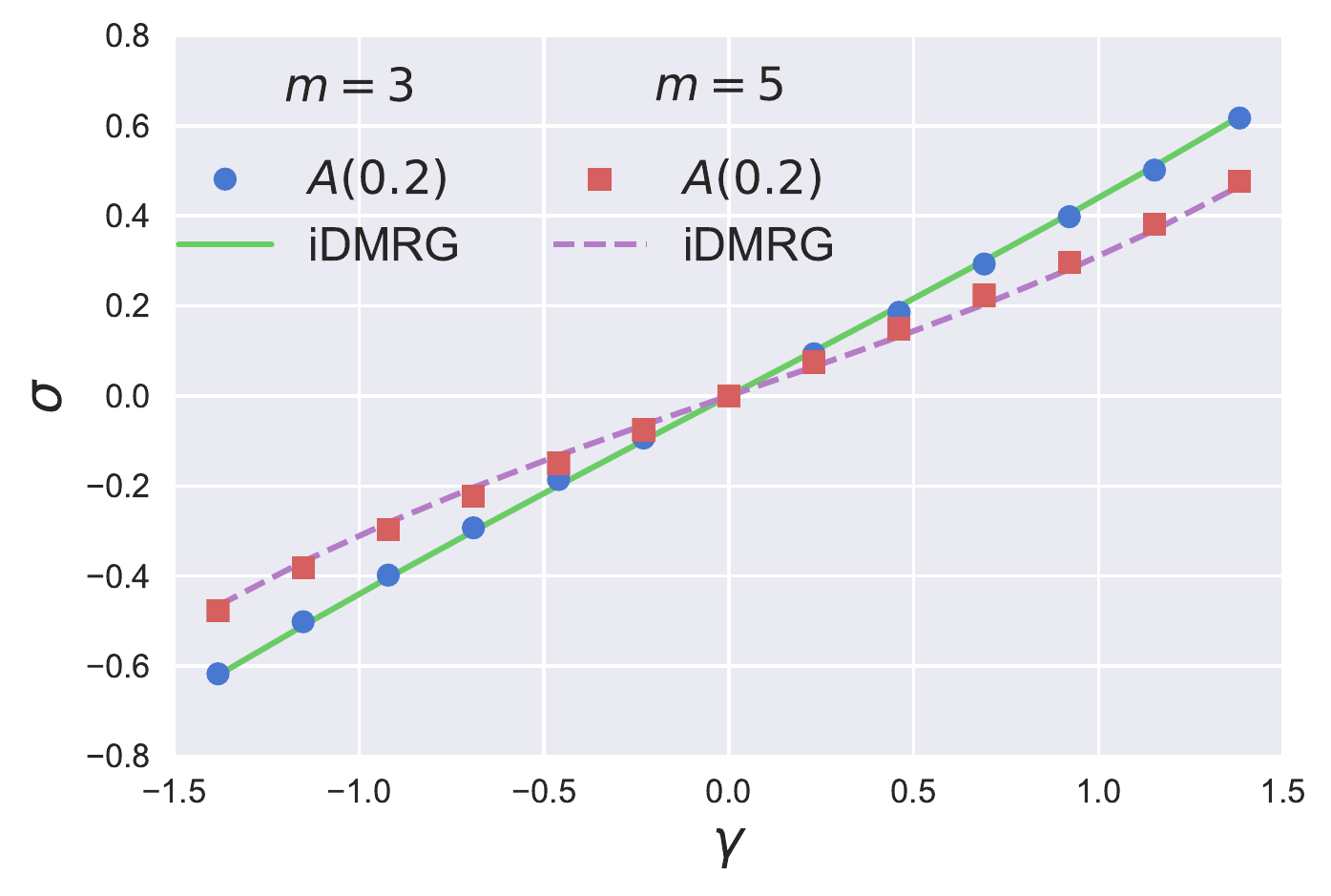}
\caption{
Comparison between the anisotropy estimate $\estimate{q=0.2}$, defined in Eq.~\eqref{eq:estimate}, and the best fit to the iDMRG data of Figures~\ref{fig:13coul} and \ref{fig:15coul} in the form $\y = c_1 \x + c_3 \x^3$.
For $m=3$ we set $c_1=0.43$, $c_3 = 0.01$. 
For $m=5$ we set $c_1 = 0.28$ and $c_3 = 0.03$.
The estimate $\estimate{0.2}$
shows good agreement with the data for both values of $m$, in the entire range $-\log 4 \leq \x \leq \log 4$.
\label{fig:comparison}
}
\end{figure}

In general, once $q$ is picked so as to best approximate the iDMRG result at certain values of anisotropy $\x$ and inverse filling $m$, 
there is no guarantee that the same estimate $\estimate{q}$ will work well at other values of $\x$ and $m$.
However, we find that the estimate $\estimate{q = 0.2}$,
{\it i.e.} the anisotropy of the equal-probability contour of $\rho_m$ from Eq.~\eqref{eq:rhom} at $1/5$ of its maximum, 
reproduces very accurately the numerical results of Section~\ref{sec:numerical} for the $\nu = 1/3$ state ($m = 3$) in a wide range of values of $\x$, as shown in Fig.~\ref{fig:comparison}.
The same estimate can then be computed for the $\nu = 1/5$ state ($m = 5$) and compared to the iDMRG data, revealing fairly accurate agreement (also shown in Fig.~\ref{fig:comparison}).
We conclude that this remarkably simple two-body model not only matches qualitative aspects of the many-body numerical data, 
but also gives a reasonable quantitative approximation.


\section{Discussion \label{sec:discussion}}

We have numerically investigated the response of incompressible fractional quantum Hall states to band mass anisotropy.
We did so by applying a new technique based on the long-wavelength limit of the guiding center structure factor, encoded in its quartic coefficient. 
This is made possible by infinite DMRG, which enables access to a continuum of momentum values in one direction and makes the determination of the quartic coefficient very accurate.

We discussed the general response of a FQH state, and in particular of its long-wavelength structure factor, to applied anisotropy.
With the symmetries of our infinite-cylinder geometry, this reduces to three parameters:
\D, describing an isotropic change in the magnitude of $S(\mb q)$;
\y, corresponding to a uniaxial squeezing;
and $\beta$, which describes the remaining $C_4$-symmetric distortion when the uniaxial squeezing is eliminated through a coordinate transformation.
We find, within numerical and finite-size accuracy, that $\beta$ vanishes and $\D$ is a constant, leaving only a unimodular metric (the uniaxial squeezing parametrized by $\y$) to characterize the geometric response of the FQH state to band mass anisotropy.

We found that states in the first Jain/hierarchy sequence appear to share the same response to band mass anisotropy, consistent with $\y \simeq {0.43}\x$ to first order near the isotropic point. 
This is distinct from the CFL response investigated in Ref.~\cite{Ippoliti2017A}, where the CFL Fermi contour anisotropy $\alpha_{\sf CF}$ was found to obey $\alpha_{\sf CF} \simeq \aE^{0.49}$, i.e. $\y \simeq 0.49\x$ in the language of the present work. 
A natural question, then, is whether the response of the CFL state is recovered as $\nu$ approaches $1/2$ along the Jain sequence $\nu  = \frac{p}{2p+1}$.
As $\nu \to 1/2$, the energy gap above the ground states decreases and finite-size effects in our numerics become more important, 
limiting the accuracy of our results and making the answer to this question more uncertain.
Nonetheless, our data on states at fillings $\nu = 1/3$, $2/5$ and $4/9$ do \emph{not} show evidence of a drift of the exponent towards the CFL value. 
Rather, they point to the interesting possibility that all states in the sequence respond in the same way.

The response of the second Jain sequence, on the other hand, is found to be radically different. 
The only state we could study in this case is $\nu = 1/5$, due to stronger finite-size effects. 
We find $\y \approx {0.28}\x$ near the isotropic point, 
a value which is clearly different from that of the first Jain sequence, 
even when larger uncertainty is taken into account.

Our numerical findings are consistent with a minimal microscopic model of flux attachment, where the geometry of the quantum Hall state is fixed by the shape of the composite boson formed by attaching $m$ ``excluded orbitals'' to each electron. 
We compute these orbitals in the relative motion problem of two electrons in the lowest Landau level and find that the shape of the composite boson correctly reproduces our numerical results as a function of electron anisotropy and filling (which is represented by the number of ``excluded orbitals'' surrounding each electron).

This microscopic model, as formulated here, only applies to Laughlin-like ``parent states'' at filling $\nu = 1/m$, and does not directly explain the observed behavior of ``daughter states'' such as $\nu = 2/5$.
Our numerical results on the sequence of states at filling $\nu = \frac{p}{2p+1}$ find compatible geometric response.
This would arise naturally in a composite fermion picture,
where all states in the sequence are interpreted as integer quantum Hall states of the same anisotropic composite fermion.
A problem with this interpretations is that the $\nu = 1/2$ CFL state may then be expected to have the same anisotropy, as well, which appears to be in contrast with numerical findings.
The development of a more complete model of flux attachment, capable of clarifying these issues, is left to future work.

\acknowledgments

We thank R. Mong and M. Zaletel for providing the DMRG libraries used in this work, and for useful discussions. 
This work was supported by Department of Energy BES Grant DE-SC0002140.

\appendix

\section{Absence of non-elliptical distortions}\label{app:check}

In this Appendix we present numerical evidence that the long-wavelength structure factor of the anisotropic $\nu = 1/3$ state is consistent with the square of a quadratic function, $S(q) \sim (g^{ab} q_a q_b)^2$. 
Physically, this implies the shape of the state is entirely characterized by an emergent metric $g^{ab}$, and the shape of long-wavelength correlations is elliptical.
In the notation of \eq{s_reparam}, this means that the coefficient $\beta$, parametrizing non-elliptical distortions, is consistent with zero within the numerical and finite-size accuracy of our method.

\begin{figure}
\centering
\includegraphics[width=0.95\columnwidth]{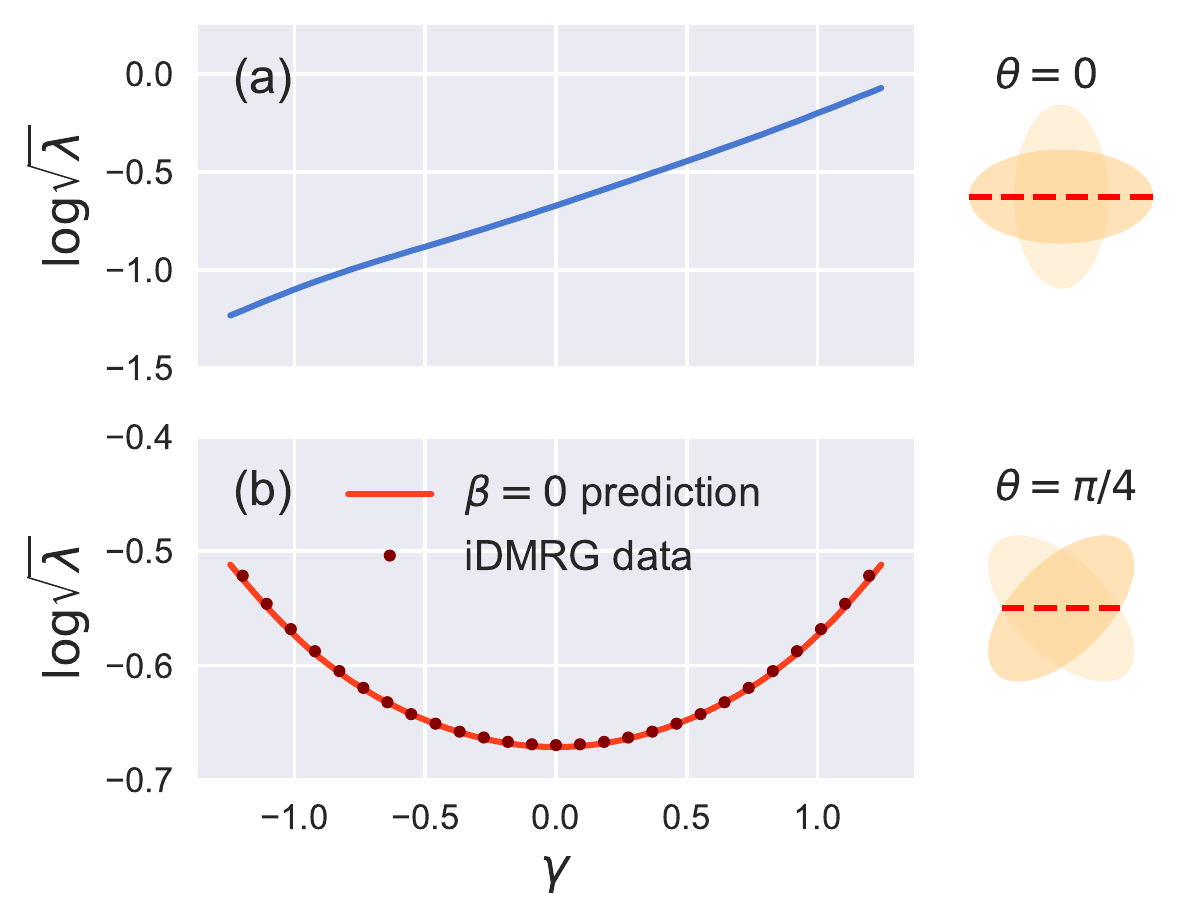}
\caption{Probing the shape of the small-$q$ structure factor by tilting the band mass tensor relative to the cylinder axis. 
(a) Numerical data for the coefficient $\lambda$ in $S(q_x, 0) \sim \lambda q_x^4$ for the $\nu = 1/3$ state as a function of the anisotropy parameter $\x$, when the electron mass tensor is aligned with the cylinder axis (ellipses shown on the right, the dashed line is the cylinder axis direction).
(b) Same data for band mass tensors tilted by $\pi/4$ relative to the cylinder axis (ellipses shown on the right).
The data agrees with a form $S(q) \sim (g^{ab}q_aq_b)^2$, obtained when the $C_4$-symmetric distortion parameter $\beta$ is set to 0.
The cylinder circumference is $L = 18\ell_B$ and the iDMRG bond dimension is $\chi = 1024$.
\label{fig:tilted}}
\end{figure}

To probe the distortions parametrized by $\beta$, we need to evaluate $S(q)$ along lines where neither $q_x$ nor $q_y$ vanish.
This is in principle possible within the setup considered in the main text, but the discretization of $q_y$ into multiples of $2\pi/L$ due to the finite circumference makes the procedure rather inaccurate.
To circumvent this issue, we consider the situation in which the band mass tensor is tilted relative to cylinder axis and circumference directions:
$$
m_{ab} = R_\theta \begin{pmatrix} e^{\x} & 0 \\ 0 & e^{-\x} \end{pmatrix} R_{-\theta}
$$
where $R_\theta$ is a rotation by an angle $\theta$.
For $\theta = \pi/4$, the general form in \eq{s_reparam} becomes
\begin{equation}
S(q_x,0)
= e^{2\D} [(\cosh \y)^2+\beta/2] q_x^4 \equiv \lambda q_x^4\;,
\label{eq:tilted_s}
\end{equation}
which allows us to directly probe $\beta$, given that $\D$ and $\y$ are known from the $\theta = 0$ measurements.

We perform iDMRG simulations of the $\nu=1/3$ state on a cylinder with circumference $L = 18\ell_B$ over a range of values of the anisotropy parameter $\x$, for mass tensors tilted by $\theta = \pi/4$ relative to the cylinder axis.
The values of $\lambda$ so obtained can be compared to the prediction of \eq{tilted_s} for $\beta = 0$, 
\begin{equation}
\sqrt \lambda = e^{\D} \cosh \y \;,
\label{eq:rotated_ansatz}
\end{equation}
to confirm the absence of non-elliptical distortions. 
The data for $\D$ and $\y$ at tilt angle $\theta = 0$ for the same filling and size was shown in Fig.~\ref{fig:13coul}(b-c), and we also show $\log \sqrt \lambda = \D + \y$ in Fig.~\ref{fig:tilted}(a) for convenience.
Fig.~\ref{fig:tilted}(b) shows iDMRG data for the $\lambda$ coefficient at tilt angle $\theta = \pi/4$, 
along with the prediction in \eq{tilted_s}, where we set $\D(\x)$ to its isotropic value $\D (0)$, 
neglecting small variations with $\x$ which have been shown to be a finite-size effect (Section~\ref{sec:13}).
As can be seen, the comparison reveals very good agreement.
We conclude that the distortion of $S(q)$ is consistent with a purely uniaxial stretching, parametrized by an internal unimodular metric whose anisotropy parameter is $\y$.

\section{Effective interaction potential in real space \label{app:veff}}

In this Appendix we derive a formula for the real-space effective interaction between two electrons in the lowest Landau level with anisotropic mass and present qualitative aspects of its shape.
This leads to a first, semi-classical approximation to the problem of anisotropic flux attachment, 
and informs the discussion in Section~\ref{sec:theory} of the main text.

We start from the effective potential for the relative coordinate problem as computed in Eq.~\eqref{eq:relative},
\begin{equation}
\tilde{V}(\mathbf q /\sqrt{2} ) 
= \frac{2^{3/2} \pi}{q} e^{-\frac{1}{4} \ell_B^2 (e^\x q_x^2 + e^{-\x} q_y^2)} \, \;.
\end{equation}
The quantity to calculate is 
\begin{equation}
\tilde{V}(x, y) =  \int d q_x dq_y \ \frac{1}{\sqrt{2} \pi q} 
e^{-\frac{1}{4} \ell_B^2 (e^\x q_x^2 + e^{-\x} q_y^2)} 
e^{i q_x x +i q_y y}
\end{equation}
We do the integral in polar coordinates $q, \theta$ and introduce shorthand notations $c \equiv \cos \theta$, $s \equiv \sin \theta$:
\begin{align}
\tilde{V}(x, y) 
& =\sqrt{2}  \int_0^{2\pi} \frac{d\theta}{2\pi} \int_0^\infty dq  \ 
e^{-\frac{\ell_B^2q^2}{4}(e^\x c^2 + e^{-\x} s^2) - iq(cx+sy)}
\nonumber \\
& = \sqrt{2} \int_0^\pi \frac{d\theta}{2\pi} \int_{-\infty}^{+\infty} dq\, 
e^{-\frac{\ell_B^2 q^2}{4}(e^\x c^2 + e^{-\x} s^2) - iq(cx+sy)}
\nonumber \\
& =  \int_0^\pi \frac{d\theta}{\pi} 
\sqrt{\frac{2 \pi \ell_B^{-2}}{e^\x c^2 + e^{-\x} s^2}}  
e^{- \ell_B^{-2} \frac{(cx+sy)^2}{e^\x c^2 + e^{-\x} s^2}}\;.
\label{eq:integral}
\end{align}
Simple manipulations and a change of integration variable $\theta \mapsto 2\theta$ lead to the expression
\begin{align}
&\tilde{V}(x, y) 
 =  \int_0^{2\pi} \frac{d\theta}{2\pi} \sqrt{\frac{2\pi \ell_B^{-2} }{\cosh \x + \cos \theta \sinh \x }}  \nonumber \\
& \quad \times \exp \left\{-\frac{1}{2} \frac{(x^2 + y^2) + (x^2-y^2)\cos\theta + 2xy\sin\theta}{\cosh \x + \cos \theta \sinh \x }\right\}\;.
\label{eq:integral}
\end{align}
The integral can be done analytically in the isotropic case $\x = 0$, 
but has to be evaluated numerically for $\x \neq 0$.

\begin{figure}
\centering
\includegraphics[width = 0.95\columnwidth]{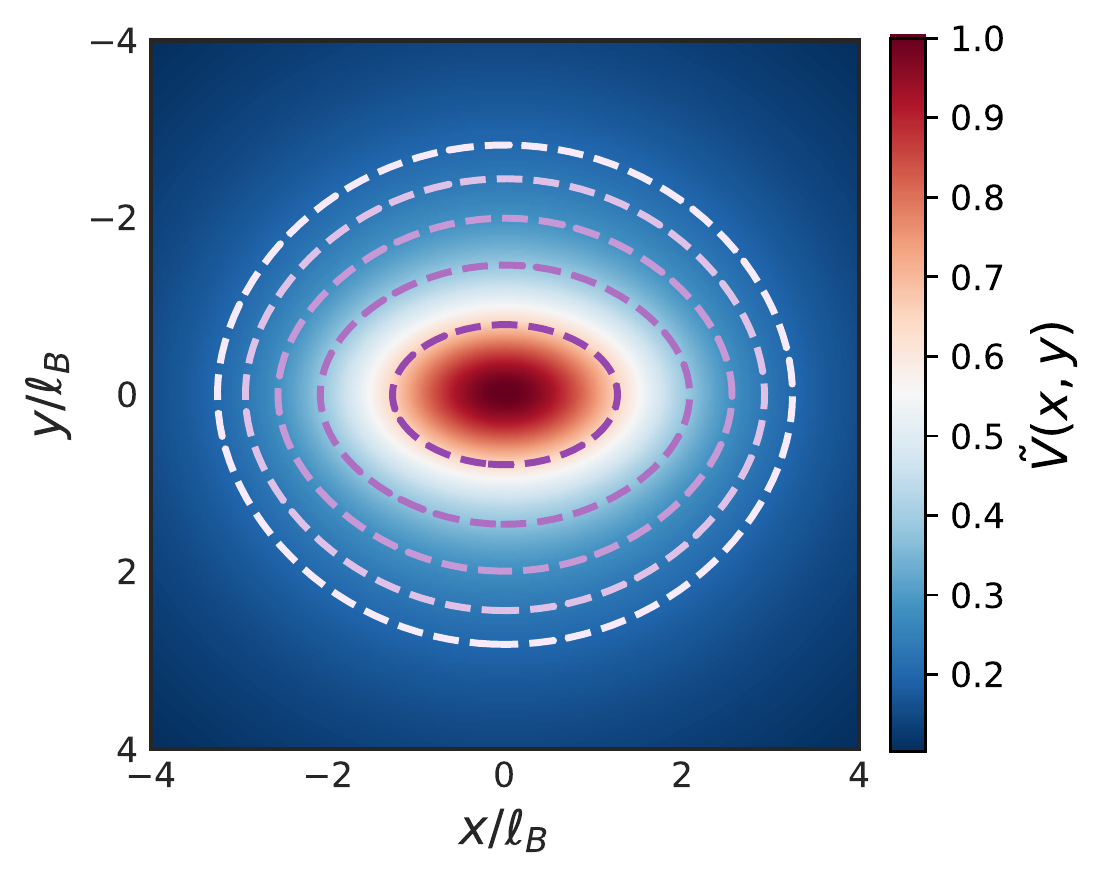}
\caption{Color plot of the real-space potential in Eq.~\eqref{eq:integral} for $\x = \log 2$.
The dashed lines are equipotential contours enclosing areas $2\pi \ell_B^2(n+1/2)$, for $n = 0$ to $4$.
\label{fig:rs_potential}}
\end{figure}

We show a plot of $\tilde{V}(x, y)$ for $\x = \log 2$ in Fig.~\ref{fig:rs_potential},
alongside equipotential contours that enclose areas $2\pi \ell_B^2(n+1/2)$, which represent an approximation of the semiclassical trajectories.
More accurately, the semiclassical orbitals should be contained between the contours enclosing areas $2\pi \ell_B^2 n$ and $2\pi\ell_B^2 (n+1)$, with $n=0$ giving a disk and $n>0$ giving annuli\cite{Haldane2016}.
The $n$-th annulus can be though of as a fattened version of the orbit enclosing area $2\pi \ell_B^2 (n+1/2)$, which is the one we consider. 

A first approximation of the anisotropy of the quantum Hall state $\y$ is then given by taking the ratio of the semiaxes of those contours.
The result underestimates $|\y|$, i.e. suggests a less anisotropic state, relative to the numerical results of  Sec.~\ref{sec:numerical}.
Nonetheless, it shows qualitatively correct behavior, with the shape of the composite boson becoming less anisotropic as $m$ (the number of attached fluxes) increases.

A more accurate approximation is given by solving the quantum mechanical problem of a LLL electron moving the potential $\tilde{V}(\mb r)$ and describing the shape of the actual excluded orbitals, rather than their semiclassical approximation.
This approach in described in the main text, Section~\ref{sec:theory}.

\section{Numerical data on $\nu = 2/5$, $4/9$ \label{app:more_numerics}}

\begin{figure}
\centering
\includegraphics[width = 0.49\textwidth]{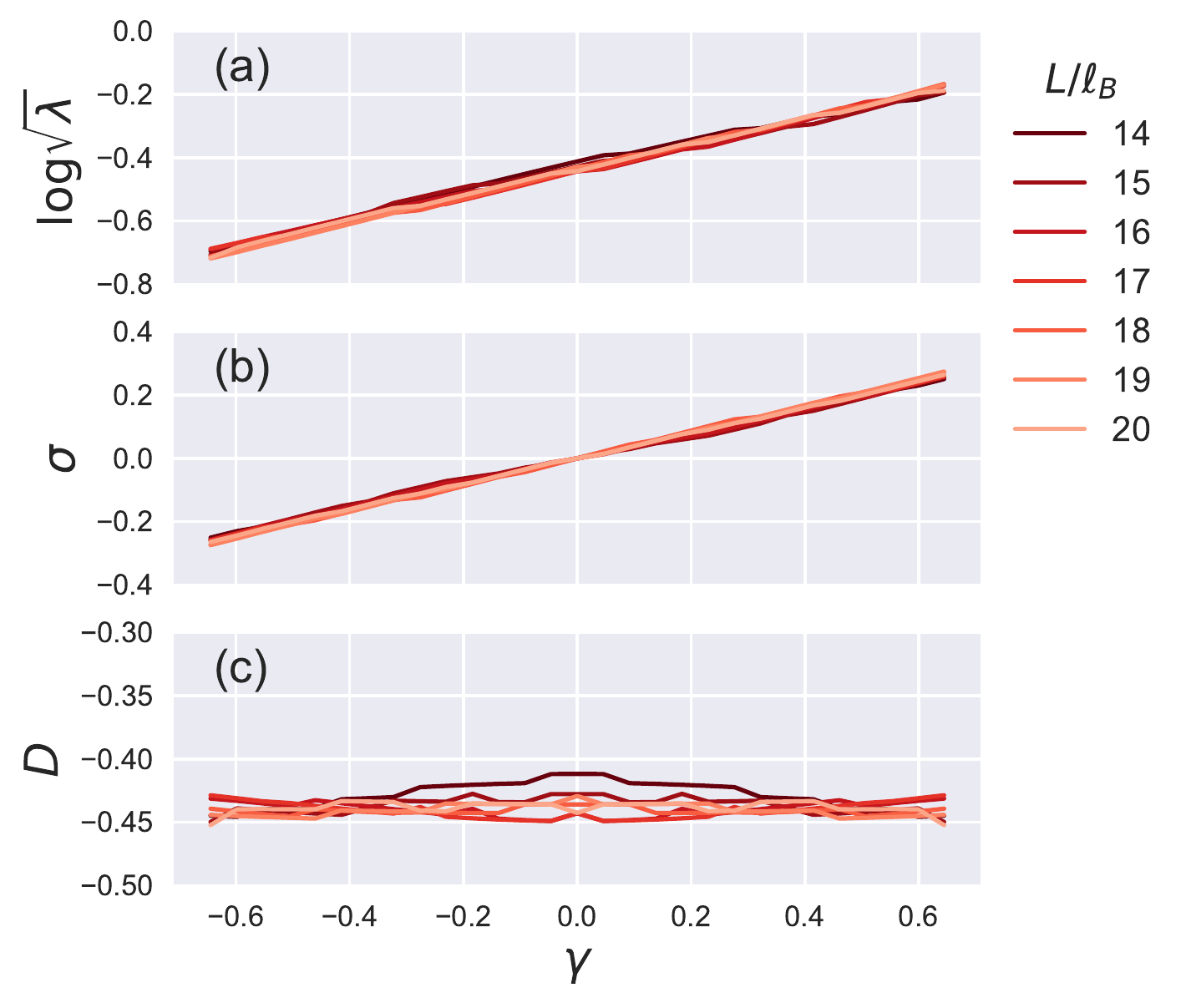} 
\caption{
Numerical data for the $\nu = 2/5$ state as a function of electron anisotropy $\x$ and cylinder circumference $L$.
The data is obtained with iDMRG at bond dimension $\chi = 1024$.
(a) Quartic coefficient $\lambda(\x)$, extracted from $S(q_x, 0) \approx \lambda(\x) q_x^4$.
(b) $\y(\x)$, defined as the odd part of $\log\sqrt{\lambda}$.
(c) $\D(\x)$, defined as the even part of $\log\sqrt{\lambda}$, has small finite-size fluctuations and approaches a constant as $L$ increases.
This data display the same qualitative behavior as the $\nu =1/3$ state (Fig.~\ref{fig:13coul}), with somewhat stronger finite-size effects.
\label{fig:25coul}
}
\end{figure}

\begin{figure}
\centering
\includegraphics[width = 0.49\textwidth]{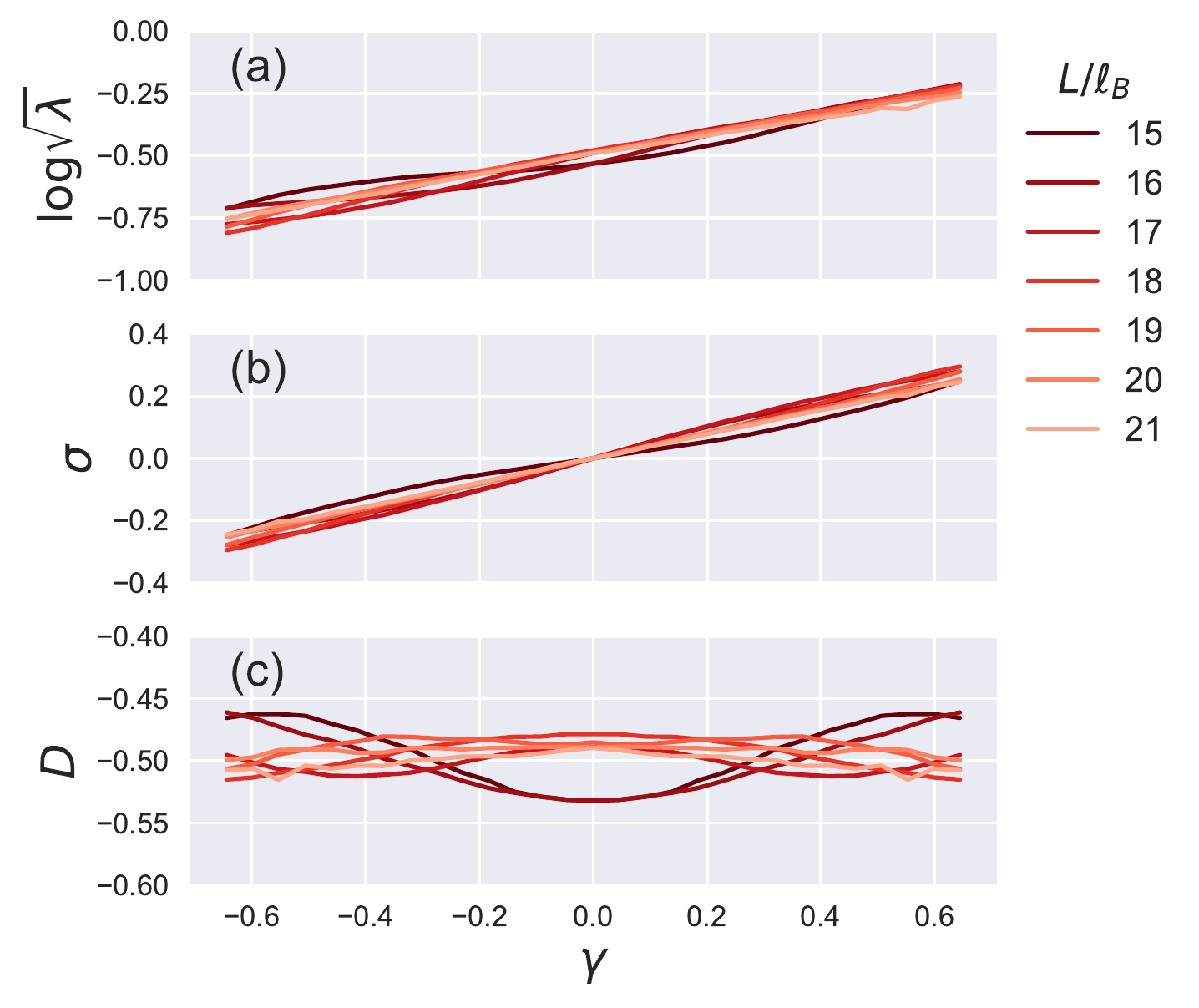}
\caption{
Same plots as those in Fig.~\ref{fig:25coul}, but for the state at filling  $\nu = 4/9$.
\label{fig:49coul}
}
\end{figure}

\begin{figure}
\centering
\includegraphics[width = 0.49\textwidth]{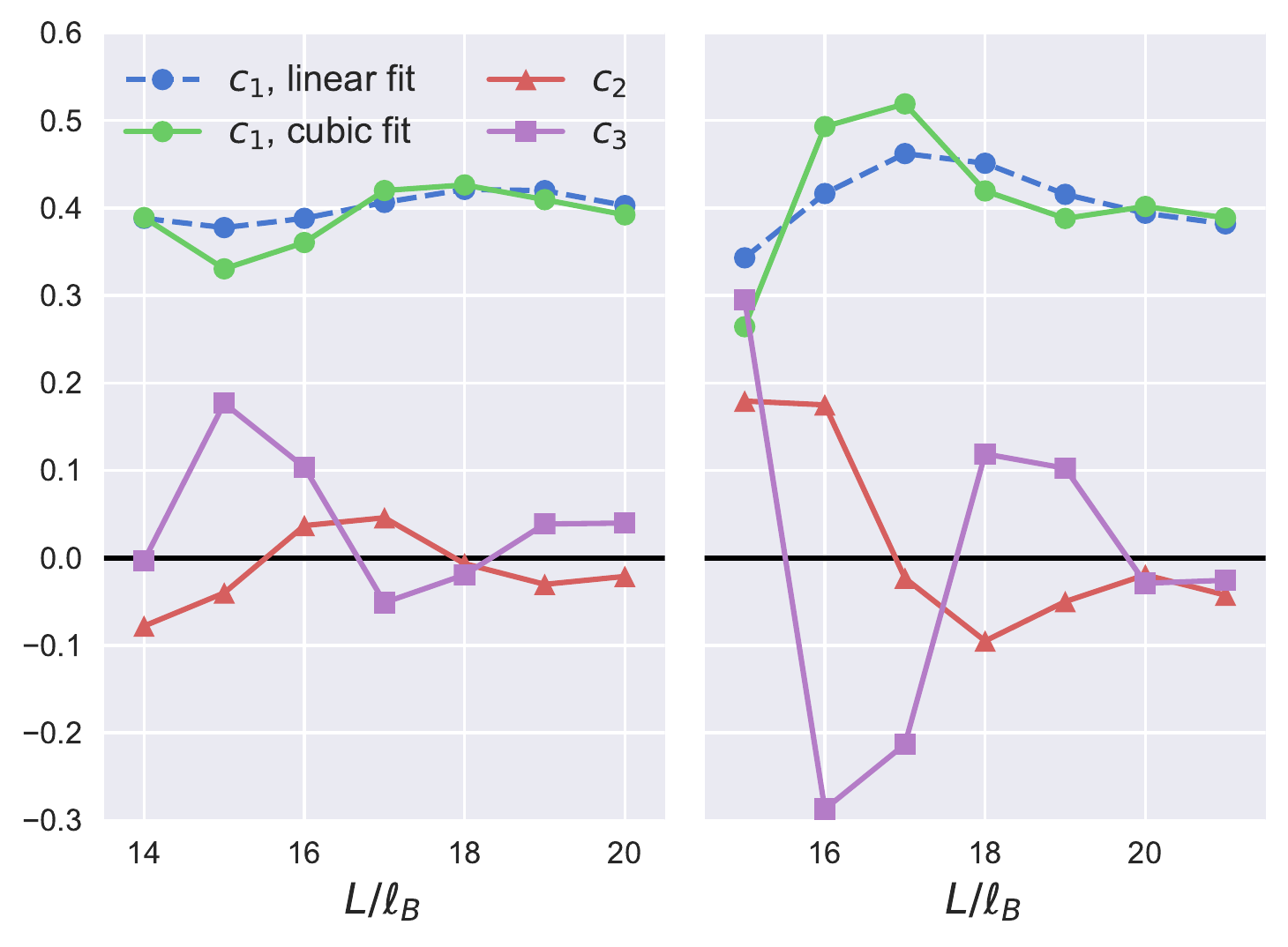}
\caption{
Coefficients of the polynomial fit of $\log\sqrt{\lambda}$ to $\x$ for the states at filling $\nu=2/5$ (left, data from Fig.~\ref{fig:25coul}) and $\nu=4/9$ (right, data from Fig.~\ref{fig:49coul}).
Despite the increasing finite-size effects, the coefficients $c_1$, $c_2$, $c_3$ are consistent with those of the $\nu = 1/3$ state, shown in Fig.~\ref{fig:coeffs13}.
\label{fig:coeffs2549}
}
\end{figure}

Here we show additional numerical data on other states in the first Jain sequence beyond $\nu = 1/3$, namely $\nu = 2/5$ and $\nu = 4/9$.
As the filling fraction approaches $\nu = 1/2$ along the sequence $\nu = \frac{p}{2p+1}$, the gap of the associated incompressible state becomes smaller and its correlation length become longer.
This implies stronger finite-size effects compared to the $\nu = 1/3$ state.

The data we obtain for the quartic coefficient $\lambda$ in $S(q_x, 0)\approx \lambda q_x^4$ is shown in Fig.~\ref{fig:25coul}(a) and \ref{fig:49coul}(a) for filling fractions $\nu = 2/5$ and $\nu = 4/9$ respectively.
The measured values of $\lambda(\x)$, despite finite-circumference fluctuations, clearly show a common trend.
This is further illustrated by splitting $\log \sqrt{\lambda}$ into its even and odd parts, $\D$ and $\y$ (Fig.~\ref{fig:25coul}(b,c) and \ref{fig:49coul}(b,c)),
and by performing a polynomial fit of $\D + \y$ as a function of $\x$, \eq{eq:polyfit}.
The fit coefficients $c_{1, 2, 3}$ are shown, for both fillings, in Fig.~\ref{fig:coeffs2549}.
They display wider fluctuations with circumference $L$ compared to the $\nu = 1/3$ state (Fig.~\ref{fig:coeffs13}), 
but appear to be consistent with the same asymptotic values, 
$c_1 \simeq 0.42$, $c_2 \simeq 0$, $c_3 \simeq 0.02$.

These data suggest that all incompressible states in the sequence $\nu = \frac{p}{2p+1}$ respond quantitatively in the same way to applied anisotropy, and the response to mass anisotropy does not appear to drift towards that observed for the CFL state in Ref.~\cite{Ippoliti2017A}, $c_1 \simeq 0.49$.

\clearpage

\bibliography{gapped_anisotropy}

\begin{thebibliography}{40}%
\makeatletter
\providecommand \@ifxundefined [1]{%
 \@ifx{#1\undefined}
}%
\providecommand \@ifnum [1]{%
 \ifnum #1\expandafter \@firstoftwo
 \else \expandafter \@secondoftwo
 \fi
}%
\providecommand \@ifx [1]{%
 \ifx #1\expandafter \@firstoftwo
 \else \expandafter \@secondoftwo
 \fi
}%
\providecommand \natexlab [1]{#1}%
\providecommand \enquote  [1]{``#1''}%
\providecommand \bibnamefont  [1]{#1}%
\providecommand \bibfnamefont [1]{#1}%
\providecommand \citenamefont [1]{#1}%
\providecommand \href@noop [0]{\@secondoftwo}%
\providecommand \href [0]{\begingroup \@sanitize@url \@href}%
\providecommand \@href[1]{\@@startlink{#1}\@@href}%
\providecommand \@@href[1]{\endgroup#1\@@endlink}%
\providecommand \@sanitize@url [0]{\catcode `\\12\catcode `\$12\catcode
  `\&12\catcode `\#12\catcode `\^12\catcode `\_12\catcode `\%12\relax}%
\providecommand \@@startlink[1]{}%
\providecommand \@@endlink[0]{}%
\providecommand \url  [0]{\begingroup\@sanitize@url \@url }%
\providecommand \@url [1]{\endgroup\@href {#1}{\urlprefix }}%
\providecommand \urlprefix  [0]{URL }%
\providecommand \Eprint [0]{\href }%
\providecommand \doibase [0]{http://dx.doi.org/}%
\providecommand \selectlanguage [0]{\@gobble}%
\providecommand \bibinfo  [0]{\@secondoftwo}%
\providecommand \bibfield  [0]{\@secondoftwo}%
\providecommand \translation [1]{[#1]}%
\providecommand \BibitemOpen [0]{}%
\providecommand \bibitemStop [0]{}%
\providecommand \bibitemNoStop [0]{.\EOS\space}%
\providecommand \EOS [0]{\spacefactor3000\relax}%
\providecommand \BibitemShut  [1]{\csname bibitem#1\endcsname}%
\let\auto@bib@innerbib\@empty
\bibitem [{\citenamefont {Haldane}(2011)}]{Haldane2011}%
  \BibitemOpen
  \bibfield  {author} {\bibinfo {author} {\bibfnamefont {F.~D.~M.}\
  \bibnamefont {Haldane}},\ }\href {\doibase 10.1103/PhysRevLett.107.116801}
  {\bibfield  {journal} {\bibinfo  {journal} {Phys. Rev. Lett.}\ }\textbf
  {\bibinfo {volume} {107}},\ \bibinfo {pages} {116801} (\bibinfo {year}
  {2011})}\BibitemShut {NoStop}%
\bibitem [{\citenamefont {Haldane}(1983{\natexlab{a}})}]{HaldanePP}%
  \BibitemOpen
  \bibfield  {author} {\bibinfo {author} {\bibfnamefont {F.~D.~M.}\
  \bibnamefont {Haldane}},\ }\href {\doibase 10.1103/PhysRevLett.51.605}
  {\bibfield  {journal} {\bibinfo  {journal} {Phys. Rev. Lett.}\ }\textbf
  {\bibinfo {volume} {51}},\ \bibinfo {pages} {605} (\bibinfo {year}
  {1983}{\natexlab{a}})}\BibitemShut {NoStop}%
\bibitem [{\citenamefont {Yang}\ \emph
  {et~al.}(2017{\natexlab{a}})\citenamefont {Yang}, \citenamefont {Hu},
  \citenamefont {Lee},\ and\ \citenamefont {Papi\ifmmode~\acute{c}\else
  \'{c}\fi{}}}]{BoYang2017A}%
  \BibitemOpen
  \bibfield  {author} {\bibinfo {author} {\bibfnamefont {B.}~\bibnamefont
  {Yang}}, \bibinfo {author} {\bibfnamefont {Z.-X.}\ \bibnamefont {Hu}},
  \bibinfo {author} {\bibfnamefont {C.~H.}\ \bibnamefont {Lee}}, \ and\
  \bibinfo {author} {\bibfnamefont {Z.}~\bibnamefont
  {Papi\ifmmode~\acute{c}\else \'{c}\fi{}}},\ }\href {\doibase
  10.1103/PhysRevLett.118.146403} {\bibfield  {journal} {\bibinfo  {journal}
  {Phys. Rev. Lett.}\ }\textbf {\bibinfo {volume} {118}},\ \bibinfo {pages}
  {146403} (\bibinfo {year} {2017}{\natexlab{a}})}\BibitemShut {NoStop}%
\bibitem [{\citenamefont {Yang}\ \emph
  {et~al.}(2017{\natexlab{b}})\citenamefont {Yang}, \citenamefont {Lee},
  \citenamefont {Zhang},\ and\ \citenamefont {Hu}}]{BoYang2017B}%
  \BibitemOpen
  \bibfield  {author} {\bibinfo {author} {\bibfnamefont {B.}~\bibnamefont
  {Yang}}, \bibinfo {author} {\bibfnamefont {C.~H.}\ \bibnamefont {Lee}},
  \bibinfo {author} {\bibfnamefont {C.}~\bibnamefont {Zhang}}, \ and\ \bibinfo
  {author} {\bibfnamefont {Z.-X.}\ \bibnamefont {Hu}},\ }\href {\doibase
  10.1103/PhysRevB.96.195140} {\bibfield  {journal} {\bibinfo  {journal} {Phys.
  Rev. B}\ }\textbf {\bibinfo {volume} {96}},\ \bibinfo {pages} {195140}
  (\bibinfo {year} {2017}{\natexlab{b}})}\BibitemShut {NoStop}%
\bibitem [{\citenamefont {Qiu}\ \emph {et~al.}(2012)\citenamefont {Qiu},
  \citenamefont {Haldane}, \citenamefont {Wan}, \citenamefont {Yang},\ and\
  \citenamefont {Yi}}]{Qiu2012}%
  \BibitemOpen
  \bibfield  {author} {\bibinfo {author} {\bibfnamefont {R.-Z.}\ \bibnamefont
  {Qiu}}, \bibinfo {author} {\bibfnamefont {F.~D.~M.}\ \bibnamefont {Haldane}},
  \bibinfo {author} {\bibfnamefont {X.}~\bibnamefont {Wan}}, \bibinfo {author}
  {\bibfnamefont {K.}~\bibnamefont {Yang}}, \ and\ \bibinfo {author}
  {\bibfnamefont {S.}~\bibnamefont {Yi}},\ }\href {\doibase
  10.1103/PhysRevB.85.115308} {\bibfield  {journal} {\bibinfo  {journal} {Phys.
  Rev. B}\ }\textbf {\bibinfo {volume} {85}},\ \bibinfo {pages} {115308}
  (\bibinfo {year} {2012})}\BibitemShut {NoStop}%
\bibitem [{\citenamefont {Balram}\ and\ \citenamefont
  {Jain}(2016)}]{Balram2016}%
  \BibitemOpen
  \bibfield  {author} {\bibinfo {author} {\bibfnamefont {A.~C.}\ \bibnamefont
  {Balram}}\ and\ \bibinfo {author} {\bibfnamefont {J.~K.}\ \bibnamefont
  {Jain}},\ }\href {\doibase 10.1103/PhysRevB.93.075121} {\bibfield  {journal}
  {\bibinfo  {journal} {Phys. Rev. B}\ }\textbf {\bibinfo {volume} {93}},\
  \bibinfo {pages} {075121} (\bibinfo {year} {2016})}\BibitemShut {NoStop}%
\bibitem [{\citenamefont {Can}\ \emph {et~al.}(2014)\citenamefont {Can},
  \citenamefont {Laskin},\ and\ \citenamefont {Wiegmann}}]{Can2014}%
  \BibitemOpen
  \bibfield  {author} {\bibinfo {author} {\bibfnamefont {T.}~\bibnamefont
  {Can}}, \bibinfo {author} {\bibfnamefont {M.}~\bibnamefont {Laskin}}, \ and\
  \bibinfo {author} {\bibfnamefont {P.}~\bibnamefont {Wiegmann}},\ }\href
  {\doibase 10.1103/PhysRevLett.113.046803} {\bibfield  {journal} {\bibinfo
  {journal} {Phys. Rev. Lett.}\ }\textbf {\bibinfo {volume} {113}},\ \bibinfo
  {pages} {046803} (\bibinfo {year} {2014})}\BibitemShut {NoStop}%
\bibitem [{\citenamefont {Can}\ \emph {et~al.}(2015)\citenamefont {Can},
  \citenamefont {Laskin},\ and\ \citenamefont {Wiegmann}}]{Can2015}%
  \BibitemOpen
  \bibfield  {author} {\bibinfo {author} {\bibfnamefont {T.}~\bibnamefont
  {Can}}, \bibinfo {author} {\bibfnamefont {M.}~\bibnamefont {Laskin}}, \ and\
  \bibinfo {author} {\bibfnamefont {P.~B.}\ \bibnamefont {Wiegmann}},\ }\href
  {\doibase https://doi.org/10.1016/j.aop.2015.02.013} {\bibfield  {journal}
  {\bibinfo  {journal} {Annals of Physics}\ }\textbf {\bibinfo {volume}
  {362}},\ \bibinfo {pages} {752 } (\bibinfo {year} {2015})}\BibitemShut
  {NoStop}%
\bibitem [{\citenamefont {Bradlyn}\ and\ \citenamefont
  {Read}(2015)}]{Bradlyn2015}%
  \BibitemOpen
  \bibfield  {author} {\bibinfo {author} {\bibfnamefont {B.}~\bibnamefont
  {Bradlyn}}\ and\ \bibinfo {author} {\bibfnamefont {N.}~\bibnamefont {Read}},\
  }\href {\doibase 10.1103/PhysRevB.91.165306} {\bibfield  {journal} {\bibinfo
  {journal} {Phys. Rev. B}\ }\textbf {\bibinfo {volume} {91}},\ \bibinfo
  {pages} {165306} (\bibinfo {year} {2015})}\BibitemShut {NoStop}%
\bibitem [{\citenamefont {Gromov}\ and\ \citenamefont
  {Son}(2017)}]{Gromov2017A}%
  \BibitemOpen
  \bibfield  {author} {\bibinfo {author} {\bibfnamefont {A.}~\bibnamefont
  {Gromov}}\ and\ \bibinfo {author} {\bibfnamefont {D.~T.}\ \bibnamefont
  {Son}},\ }\href {\doibase 10.1103/PhysRevX.7.041032} {\bibfield  {journal}
  {\bibinfo  {journal} {Phys. Rev. X}\ }\textbf {\bibinfo {volume} {7}},\
  \bibinfo {pages} {041032} (\bibinfo {year} {2017})}\BibitemShut {NoStop}%
\bibitem [{\citenamefont {Gromov}\ \emph {et~al.}(2017)\citenamefont {Gromov},
  \citenamefont {Geraedts},\ and\ \citenamefont {Bradlyn}}]{Gromov2017B}%
  \BibitemOpen
  \bibfield  {author} {\bibinfo {author} {\bibfnamefont {A.}~\bibnamefont
  {Gromov}}, \bibinfo {author} {\bibfnamefont {S.~D.}\ \bibnamefont
  {Geraedts}}, \ and\ \bibinfo {author} {\bibfnamefont {B.}~\bibnamefont
  {Bradlyn}},\ }\href {\doibase 10.1103/PhysRevLett.119.146602} {\bibfield
  {journal} {\bibinfo  {journal} {Phys. Rev. Lett.}\ }\textbf {\bibinfo
  {volume} {119}},\ \bibinfo {pages} {146602} (\bibinfo {year}
  {2017})}\BibitemShut {NoStop}%
\bibitem [{\citenamefont {You}\ \emph {et~al.}(2014)\citenamefont {You},
  \citenamefont {Cho},\ and\ \citenamefont {Fradkin}}]{You2014}%
  \BibitemOpen
  \bibfield  {author} {\bibinfo {author} {\bibfnamefont {Y.}~\bibnamefont
  {You}}, \bibinfo {author} {\bibfnamefont {G.~Y.}\ \bibnamefont {Cho}}, \ and\
  \bibinfo {author} {\bibfnamefont {E.}~\bibnamefont {Fradkin}},\ }\href
  {\doibase 10.1103/PhysRevX.4.041050} {\bibfield  {journal} {\bibinfo
  {journal} {Phys. Rev. X}\ }\textbf {\bibinfo {volume} {4}},\ \bibinfo {pages}
  {041050} (\bibinfo {year} {2014})}\BibitemShut {NoStop}%
\bibitem [{\citenamefont {Zhu}\ \emph {et~al.}(2017)\citenamefont {Zhu},
  \citenamefont {Sodemann}, \citenamefont {Sheng},\ and\ \citenamefont
  {Fu}}]{Zhu2017}%
  \BibitemOpen
  \bibfield  {author} {\bibinfo {author} {\bibfnamefont {Z.}~\bibnamefont
  {Zhu}}, \bibinfo {author} {\bibfnamefont {I.}~\bibnamefont {Sodemann}},
  \bibinfo {author} {\bibfnamefont {D.~N.}\ \bibnamefont {Sheng}}, \ and\
  \bibinfo {author} {\bibfnamefont {L.}~\bibnamefont {Fu}},\ }\href {\doibase
  10.1103/PhysRevB.95.201116} {\bibfield  {journal} {\bibinfo  {journal} {Phys.
  Rev. B}\ }\textbf {\bibinfo {volume} {95}},\ \bibinfo {pages} {201116}
  (\bibinfo {year} {2017})}\BibitemShut {NoStop}%
\bibitem [{\citenamefont {Zhu}\ \emph {et~al.}(2018)\citenamefont {Zhu},
  \citenamefont {Sheng}, \citenamefont {Fu},\ and\ \citenamefont
  {Sodemann}}]{Zhu2018}%
  \BibitemOpen
  \bibfield  {author} {\bibinfo {author} {\bibfnamefont {Z.}~\bibnamefont
  {Zhu}}, \bibinfo {author} {\bibfnamefont {D.~N.}\ \bibnamefont {Sheng}},
  \bibinfo {author} {\bibfnamefont {L.}~\bibnamefont {Fu}}, \ and\ \bibinfo
  {author} {\bibfnamefont {I.}~\bibnamefont {Sodemann}},\ }\href@noop {}
  {\bibfield  {journal} {\bibinfo  {journal} {arXiv}\ ,\ \bibinfo {pages}
  {1802.02167}} (\bibinfo {year} {2018})}\BibitemShut {NoStop}%
\bibitem [{\citenamefont {Lee}\ \emph {et~al.}(2018)\citenamefont {Lee},
  \citenamefont {Shao}, \citenamefont {Kim}, \citenamefont {Haldane},\ and\
  \citenamefont {Rezayi}}]{Lee2018}%
  \BibitemOpen
  \bibfield  {author} {\bibinfo {author} {\bibfnamefont {K.}~\bibnamefont
  {Lee}}, \bibinfo {author} {\bibfnamefont {J.}~\bibnamefont {Shao}}, \bibinfo
  {author} {\bibfnamefont {E.-A.}\ \bibnamefont {Kim}}, \bibinfo {author}
  {\bibfnamefont {F.~D.~M.}\ \bibnamefont {Haldane}}, \ and\ \bibinfo {author}
  {\bibfnamefont {E.}~\bibnamefont {Rezayi}},\ }\href@noop {} {\bibfield
  {journal} {\bibinfo  {journal} {arXiv}\ ,\ \bibinfo {pages} {1802.08261}}
  (\bibinfo {year} {2018})}\BibitemShut {NoStop}%
\bibitem [{\citenamefont {Nguyen}\ \emph {et~al.}(2018)\citenamefont {Nguyen},
  \citenamefont {Gromov},\ and\ \citenamefont {Son}}]{Nguyen2017}%
  \BibitemOpen
  \bibfield  {author} {\bibinfo {author} {\bibfnamefont {D.~X.}\ \bibnamefont
  {Nguyen}}, \bibinfo {author} {\bibfnamefont {A.}~\bibnamefont {Gromov}}, \
  and\ \bibinfo {author} {\bibfnamefont {D.~T.}\ \bibnamefont {Son}},\ }\href
  {\doibase 10.1103/PhysRevB.97.195103} {\bibfield  {journal} {\bibinfo
  {journal} {Phys. Rev. B}\ }\textbf {\bibinfo {volume} {97}},\ \bibinfo
  {pages} {195103} (\bibinfo {year} {2018})}\BibitemShut {NoStop}%
\bibitem [{\citenamefont {Ippoliti}\ \emph
  {et~al.}(2017{\natexlab{a}})\citenamefont {Ippoliti}, \citenamefont
  {Geraedts},\ and\ \citenamefont {Bhatt}}]{Ippoliti2017C}%
  \BibitemOpen
  \bibfield  {author} {\bibinfo {author} {\bibfnamefont {M.}~\bibnamefont
  {Ippoliti}}, \bibinfo {author} {\bibfnamefont {S.~D.}\ \bibnamefont
  {Geraedts}}, \ and\ \bibinfo {author} {\bibfnamefont {R.~N.}\ \bibnamefont
  {Bhatt}},\ }\href {\doibase 10.1103/PhysRevB.96.115151} {\bibfield  {journal}
  {\bibinfo  {journal} {Phys. Rev. B}\ }\textbf {\bibinfo {volume} {96}},\
  \bibinfo {pages} {115151} (\bibinfo {year} {2017}{\natexlab{a}})}\BibitemShut
  {NoStop}%
\bibitem [{\citenamefont {Liu}\ \emph {et~al.}(2018)\citenamefont {Liu},
  \citenamefont {Gromov},\ and\ \citenamefont {Papic}}]{Liu2018}%
  \BibitemOpen
  \bibfield  {author} {\bibinfo {author} {\bibfnamefont {Z.}~\bibnamefont
  {Liu}}, \bibinfo {author} {\bibfnamefont {A.}~\bibnamefont {Gromov}}, \ and\
  \bibinfo {author} {\bibfnamefont {Z.}~\bibnamefont {Papic}},\ }\href@noop {}
  {\bibfield  {journal} {\bibinfo  {journal} {arXiv}\ ,\ \bibinfo {pages}
  {1803.00030}} (\bibinfo {year} {2018})}\BibitemShut {NoStop}%
\bibitem [{\citenamefont {Gokmen}\ \emph {et~al.}(2010)\citenamefont {Gokmen},
  \citenamefont {Padmanabhan},\ and\ \citenamefont {Shayegan}}]{Gokmen2010}%
  \BibitemOpen
  \bibfield  {author} {\bibinfo {author} {\bibfnamefont {T.}~\bibnamefont
  {Gokmen}}, \bibinfo {author} {\bibfnamefont {M.}~\bibnamefont {Padmanabhan}},
  \ and\ \bibinfo {author} {\bibfnamefont {M.}~\bibnamefont {Shayegan}},\
  }\href@noop {} {\bibfield  {journal} {\bibinfo  {journal} {Nat. Phys.}\
  }\textbf {\bibinfo {volume} {6}},\ \bibinfo {pages} {621} (\bibinfo {year}
  {2010})}\BibitemShut {NoStop}%
\bibitem [{\citenamefont {Kamburov}\ \emph {et~al.}(2013)\citenamefont
  {Kamburov}, \citenamefont {Liu}, \citenamefont {Shayegan}, \citenamefont
  {Pfeiffer}, \citenamefont {West},\ and\ \citenamefont
  {Baldwin}}]{Kamburov2013}%
  \BibitemOpen
  \bibfield  {author} {\bibinfo {author} {\bibfnamefont {D.}~\bibnamefont
  {Kamburov}}, \bibinfo {author} {\bibfnamefont {Y.}~\bibnamefont {Liu}},
  \bibinfo {author} {\bibfnamefont {M.}~\bibnamefont {Shayegan}}, \bibinfo
  {author} {\bibfnamefont {L.~N.}\ \bibnamefont {Pfeiffer}}, \bibinfo {author}
  {\bibfnamefont {K.~W.}\ \bibnamefont {West}}, \ and\ \bibinfo {author}
  {\bibfnamefont {K.~W.}\ \bibnamefont {Baldwin}},\ }\href {\doibase
  10.1103/PhysRevLett.110.206801} {\bibfield  {journal} {\bibinfo  {journal}
  {Phys. Rev. Lett.}\ }\textbf {\bibinfo {volume} {110}},\ \bibinfo {pages}
  {206801} (\bibinfo {year} {2013})}\BibitemShut {NoStop}%
\bibitem [{\citenamefont {Jo}\ \emph {et~al.}(2017)\citenamefont {Jo},
  \citenamefont {Rosales}, \citenamefont {Mueed}, \citenamefont {Pfeiffer},
  \citenamefont {West}, \citenamefont {Baldwin}, \citenamefont {Winkler},
  \citenamefont {Padmanabhan},\ and\ \citenamefont {Shayegan}}]{Jo2017}%
  \BibitemOpen
  \bibfield  {author} {\bibinfo {author} {\bibfnamefont {I.}~\bibnamefont
  {Jo}}, \bibinfo {author} {\bibfnamefont {K.~A.~V.}\ \bibnamefont {Rosales}},
  \bibinfo {author} {\bibfnamefont {M.~A.}\ \bibnamefont {Mueed}}, \bibinfo
  {author} {\bibfnamefont {L.~N.}\ \bibnamefont {Pfeiffer}}, \bibinfo {author}
  {\bibfnamefont {K.~W.}\ \bibnamefont {West}}, \bibinfo {author}
  {\bibfnamefont {K.~W.}\ \bibnamefont {Baldwin}}, \bibinfo {author}
  {\bibfnamefont {R.}~\bibnamefont {Winkler}}, \bibinfo {author} {\bibfnamefont
  {M.}~\bibnamefont {Padmanabhan}}, \ and\ \bibinfo {author} {\bibfnamefont
  {M.}~\bibnamefont {Shayegan}},\ }\href {\doibase
  10.1103/PhysRevLett.119.016402} {\bibfield  {journal} {\bibinfo  {journal}
  {Phys. Rev. Lett.}\ }\textbf {\bibinfo {volume} {119}},\ \bibinfo {pages}
  {016402} (\bibinfo {year} {2017})}\BibitemShut {NoStop}%
\bibitem [{\citenamefont {Mueed}\ \emph {et~al.}(2015)\citenamefont {Mueed},
  \citenamefont {Kamburov}, \citenamefont {Liu}, \citenamefont {Shayegan},
  \citenamefont {Pfeiffer}, \citenamefont {West}, \citenamefont {Baldwin},\
  and\ \citenamefont {Winkler}}]{Mueed2015}%
  \BibitemOpen
  \bibfield  {author} {\bibinfo {author} {\bibfnamefont {M.~A.}\ \bibnamefont
  {Mueed}}, \bibinfo {author} {\bibfnamefont {D.}~\bibnamefont {Kamburov}},
  \bibinfo {author} {\bibfnamefont {Y.}~\bibnamefont {Liu}}, \bibinfo {author}
  {\bibfnamefont {M.}~\bibnamefont {Shayegan}}, \bibinfo {author}
  {\bibfnamefont {L.~N.}\ \bibnamefont {Pfeiffer}}, \bibinfo {author}
  {\bibfnamefont {K.~W.}\ \bibnamefont {West}}, \bibinfo {author}
  {\bibfnamefont {K.~W.}\ \bibnamefont {Baldwin}}, \ and\ \bibinfo {author}
  {\bibfnamefont {R.}~\bibnamefont {Winkler}},\ }\href {\doibase
  10.1103/PhysRevLett.114.176805} {\bibfield  {journal} {\bibinfo  {journal}
  {Phys. Rev. Lett.}\ }\textbf {\bibinfo {volume} {114}},\ \bibinfo {pages}
  {176805} (\bibinfo {year} {2015})}\BibitemShut {NoStop}%
\bibitem [{\citenamefont {Kamburov}\ \emph {et~al.}(2014)\citenamefont
  {Kamburov}, \citenamefont {Mueed}, \citenamefont {Shayegan}, \citenamefont
  {Pfeiffer}, \citenamefont {West}, \citenamefont {Baldwin}, \citenamefont
  {Lee},\ and\ \citenamefont {Winkler}}]{Kamburov2014}%
  \BibitemOpen
  \bibfield  {author} {\bibinfo {author} {\bibfnamefont {D.}~\bibnamefont
  {Kamburov}}, \bibinfo {author} {\bibfnamefont {M.~A.}\ \bibnamefont {Mueed}},
  \bibinfo {author} {\bibfnamefont {M.}~\bibnamefont {Shayegan}}, \bibinfo
  {author} {\bibfnamefont {L.~N.}\ \bibnamefont {Pfeiffer}}, \bibinfo {author}
  {\bibfnamefont {K.~W.}\ \bibnamefont {West}}, \bibinfo {author}
  {\bibfnamefont {K.~W.}\ \bibnamefont {Baldwin}}, \bibinfo {author}
  {\bibfnamefont {J.~J.~D.}\ \bibnamefont {Lee}}, \ and\ \bibinfo {author}
  {\bibfnamefont {R.}~\bibnamefont {Winkler}},\ }\href {\doibase
  10.1103/PhysRevB.89.085304} {\bibfield  {journal} {\bibinfo  {journal} {Phys.
  Rev. B}\ }\textbf {\bibinfo {volume} {89}},\ \bibinfo {pages} {085304}
  (\bibinfo {year} {2014})}\BibitemShut {NoStop}%
\bibitem [{\citenamefont {Ippoliti}\ \emph
  {et~al.}(2017{\natexlab{b}})\citenamefont {Ippoliti}, \citenamefont
  {Geraedts},\ and\ \citenamefont {Bhatt}}]{Ippoliti2017A}%
  \BibitemOpen
  \bibfield  {author} {\bibinfo {author} {\bibfnamefont {M.}~\bibnamefont
  {Ippoliti}}, \bibinfo {author} {\bibfnamefont {S.~D.}\ \bibnamefont
  {Geraedts}}, \ and\ \bibinfo {author} {\bibfnamefont {R.~N.}\ \bibnamefont
  {Bhatt}},\ }\href {\doibase 10.1103/PhysRevB.95.201104} {\bibfield  {journal}
  {\bibinfo  {journal} {Phys. Rev. B}\ }\textbf {\bibinfo {volume} {95}},\
  \bibinfo {pages} {201104(R)} (\bibinfo {year}
  {2017}{\natexlab{b}})}\BibitemShut {NoStop}%
\bibitem [{\citenamefont {Ippoliti}\ \emph
  {et~al.}(2017{\natexlab{c}})\citenamefont {Ippoliti}, \citenamefont
  {Geraedts},\ and\ \citenamefont {Bhatt}}]{Ippoliti2017B}%
  \BibitemOpen
  \bibfield  {author} {\bibinfo {author} {\bibfnamefont {M.}~\bibnamefont
  {Ippoliti}}, \bibinfo {author} {\bibfnamefont {S.~D.}\ \bibnamefont
  {Geraedts}}, \ and\ \bibinfo {author} {\bibfnamefont {R.~N.}\ \bibnamefont
  {Bhatt}},\ }\href {\doibase 10.1103/PhysRevB.96.045145} {\bibfield  {journal}
  {\bibinfo  {journal} {Phys. Rev. B}\ }\textbf {\bibinfo {volume} {96}},\
  \bibinfo {pages} {045145} (\bibinfo {year} {2017}{\natexlab{c}})}\BibitemShut
  {NoStop}%
\bibitem [{\citenamefont {Samkharadze}\ \emph {et~al.}(2016)\citenamefont
  {Samkharadze}, \citenamefont {Schrieber}, \citenamefont {Gardner},
  \citenamefont {Manfra}, \citenamefont {Fradkin},\ and\ \citenamefont
  {Csathy}}]{Samkharadze2016}%
  \BibitemOpen
  \bibfield  {author} {\bibinfo {author} {\bibfnamefont {N.}~\bibnamefont
  {Samkharadze}}, \bibinfo {author} {\bibfnamefont {K.~A.}\ \bibnamefont
  {Schrieber}}, \bibinfo {author} {\bibfnamefont {G.~C.}\ \bibnamefont
  {Gardner}}, \bibinfo {author} {\bibfnamefont {M.~J.}\ \bibnamefont {Manfra}},
  \bibinfo {author} {\bibfnamefont {E.}~\bibnamefont {Fradkin}}, \ and\
  \bibinfo {author} {\bibfnamefont {G.~A.}\ \bibnamefont {Csathy}},\
  }\href@noop {} {\bibfield  {journal} {\bibinfo  {journal} {Nat. Phys.}\
  }\textbf {\bibinfo {volume} {12}},\ \bibinfo {pages} {191} (\bibinfo {year}
  {2016})}\BibitemShut {NoStop}%
\bibitem [{\citenamefont {Yang}(2016)}]{Yang2016}%
  \BibitemOpen
  \bibfield  {author} {\bibinfo {author} {\bibfnamefont {K.}~\bibnamefont
  {Yang}},\ }\href {\doibase 10.1103/PhysRevB.93.161302} {\bibfield  {journal}
  {\bibinfo  {journal} {Phys. Rev. B}\ }\textbf {\bibinfo {volume} {93}},\
  \bibinfo {pages} {161302} (\bibinfo {year} {2016})}\BibitemShut {NoStop}%
\bibitem [{\citenamefont {Wang}\ \emph {et~al.}(2012)\citenamefont {Wang},
  \citenamefont {Narayanan}, \citenamefont {Wan},\ and\ \citenamefont
  {Zhang}}]{Wang2012}%
  \BibitemOpen
  \bibfield  {author} {\bibinfo {author} {\bibfnamefont {H.}~\bibnamefont
  {Wang}}, \bibinfo {author} {\bibfnamefont {R.}~\bibnamefont {Narayanan}},
  \bibinfo {author} {\bibfnamefont {X.}~\bibnamefont {Wan}}, \ and\ \bibinfo
  {author} {\bibfnamefont {F.}~\bibnamefont {Zhang}},\ }\href {\doibase
  10.1103/PhysRevB.86.035122} {\bibfield  {journal} {\bibinfo  {journal} {Phys.
  Rev. B}\ }\textbf {\bibinfo {volume} {86}},\ \bibinfo {pages} {035122}
  (\bibinfo {year} {2012})}\BibitemShut {NoStop}%
\bibitem [{\citenamefont {Yang}\ \emph {et~al.}(2012)\citenamefont {Yang},
  \citenamefont {Papi\ifmmode~\acute{c}\else \'{c}\fi{}}, \citenamefont
  {Rezayi}, \citenamefont {Bhatt},\ and\ \citenamefont {Haldane}}]{BoYang2012}%
  \BibitemOpen
  \bibfield  {author} {\bibinfo {author} {\bibfnamefont {B.}~\bibnamefont
  {Yang}}, \bibinfo {author} {\bibfnamefont {Z.}~\bibnamefont
  {Papi\ifmmode~\acute{c}\else \'{c}\fi{}}}, \bibinfo {author} {\bibfnamefont
  {E.~H.}\ \bibnamefont {Rezayi}}, \bibinfo {author} {\bibfnamefont {R.~N.}\
  \bibnamefont {Bhatt}}, \ and\ \bibinfo {author} {\bibfnamefont {F.~D.~M.}\
  \bibnamefont {Haldane}},\ }\href {\doibase 10.1103/PhysRevB.85.165318}
  {\bibfield  {journal} {\bibinfo  {journal} {Phys. Rev. B}\ }\textbf {\bibinfo
  {volume} {85}},\ \bibinfo {pages} {165318} (\bibinfo {year}
  {2012})}\BibitemShut {NoStop}%
\bibitem [{\citenamefont {Papi\ifmmode~\acute{c}\else
  \'{c}\fi{}}(2013)}]{Papic2013}%
  \BibitemOpen
  \bibfield  {author} {\bibinfo {author} {\bibfnamefont {Z.}~\bibnamefont
  {Papi\ifmmode~\acute{c}\else \'{c}\fi{}}},\ }\href {\doibase
  10.1103/PhysRevB.87.245315} {\bibfield  {journal} {\bibinfo  {journal} {Phys.
  Rev. B}\ }\textbf {\bibinfo {volume} {87}},\ \bibinfo {pages} {245315}
  (\bibinfo {year} {2013})}\BibitemShut {NoStop}%
\bibitem [{\citenamefont {Johri}\ \emph {et~al.}(2016)\citenamefont {Johri},
  \citenamefont {Papic}, \citenamefont {Schmitteckert}, \citenamefont {Bhatt},\
  and\ \citenamefont {Haldane}}]{Johri2016}%
  \BibitemOpen
  \bibfield  {author} {\bibinfo {author} {\bibfnamefont {S.}~\bibnamefont
  {Johri}}, \bibinfo {author} {\bibfnamefont {Z.}~\bibnamefont {Papic}},
  \bibinfo {author} {\bibfnamefont {P.}~\bibnamefont {Schmitteckert}}, \bibinfo
  {author} {\bibfnamefont {R.~N.}\ \bibnamefont {Bhatt}}, \ and\ \bibinfo
  {author} {\bibfnamefont {F.~D.~M.}\ \bibnamefont {Haldane}},\ }\href
  {http://stacks.iop.org/1367-2630/18/i=2/a=025011} {\bibfield  {journal}
  {\bibinfo  {journal} {New Journal of Physics}\ }\textbf {\bibinfo {volume}
  {18}},\ \bibinfo {pages} {025011} (\bibinfo {year} {2016})}\BibitemShut
  {NoStop}%
\bibitem [{\citenamefont {Haldane}(2009)}]{Haldane2009}%
  \BibitemOpen
  \bibfield  {author} {\bibinfo {author} {\bibfnamefont {F.~D.~M.}\
  \bibnamefont {Haldane}},\ }\href@noop {} {\bibfield  {journal} {\bibinfo
  {journal} {arXiv}\ ,\ \bibinfo {pages} {0906.1854}} (\bibinfo {year}
  {2009})}\BibitemShut {NoStop}%
\bibitem [{\citenamefont {Zaletel}\ \emph {et~al.}(2013)\citenamefont
  {Zaletel}, \citenamefont {Mong},\ and\ \citenamefont
  {Pollmann}}]{Zaletel2013}%
  \BibitemOpen
  \bibfield  {author} {\bibinfo {author} {\bibfnamefont {M.~P.}\ \bibnamefont
  {Zaletel}}, \bibinfo {author} {\bibfnamefont {R.~S.~K.}\ \bibnamefont
  {Mong}}, \ and\ \bibinfo {author} {\bibfnamefont {F.}~\bibnamefont
  {Pollmann}},\ }\href {\doibase 10.1103/PhysRevLett.110.236801} {\bibfield
  {journal} {\bibinfo  {journal} {Phys. Rev. Lett.}\ }\textbf {\bibinfo
  {volume} {110}},\ \bibinfo {pages} {236801} (\bibinfo {year}
  {2013})}\BibitemShut {NoStop}%
\bibitem [{\citenamefont {Zaletel}\ \emph {et~al.}(2015)\citenamefont
  {Zaletel}, \citenamefont {Mong}, \citenamefont {Pollmann},\ and\
  \citenamefont {Rezayi}}]{Zaletel2015}%
  \BibitemOpen
  \bibfield  {author} {\bibinfo {author} {\bibfnamefont {M.~P.}\ \bibnamefont
  {Zaletel}}, \bibinfo {author} {\bibfnamefont {R.~S.~K.}\ \bibnamefont
  {Mong}}, \bibinfo {author} {\bibfnamefont {F.}~\bibnamefont {Pollmann}}, \
  and\ \bibinfo {author} {\bibfnamefont {E.~H.}\ \bibnamefont {Rezayi}},\
  }\href {\doibase 10.1103/PhysRevB.91.045115} {\bibfield  {journal} {\bibinfo
  {journal} {Phys. Rev. B}\ }\textbf {\bibinfo {volume} {91}},\ \bibinfo
  {pages} {045115} (\bibinfo {year} {2015})}\BibitemShut {NoStop}%
\bibitem [{\citenamefont {Geraedts}\ \emph {et~al.}(2016)\citenamefont
  {Geraedts}, \citenamefont {Zaletel}, \citenamefont {Mong}, \citenamefont
  {Metlitski}, \citenamefont {Vishwanath},\ and\ \citenamefont
  {Motrunich}}]{Geraedts2016}%
  \BibitemOpen
  \bibfield  {author} {\bibinfo {author} {\bibfnamefont {S.~D.}\ \bibnamefont
  {Geraedts}}, \bibinfo {author} {\bibfnamefont {M.~P.}\ \bibnamefont
  {Zaletel}}, \bibinfo {author} {\bibfnamefont {R.~S.~K.}\ \bibnamefont
  {Mong}}, \bibinfo {author} {\bibfnamefont {M.~A.}\ \bibnamefont {Metlitski}},
  \bibinfo {author} {\bibfnamefont {A.}~\bibnamefont {Vishwanath}}, \ and\
  \bibinfo {author} {\bibfnamefont {O.~I.}\ \bibnamefont {Motrunich}},\ }\href
  {\doibase 10.1126/science.aad4302} {\bibfield  {journal} {\bibinfo  {journal}
  {Science}\ }\textbf {\bibinfo {volume} {352}},\ \bibinfo {pages} {197}
  (\bibinfo {year} {2016})}\BibitemShut {NoStop}%
\bibitem [{\citenamefont {Yang}(2013)}]{Yang2013}%
  \BibitemOpen
  \bibfield  {author} {\bibinfo {author} {\bibfnamefont {K.}~\bibnamefont
  {Yang}},\ }\href {\doibase 10.1103/PhysRevB.88.241105} {\bibfield  {journal}
  {\bibinfo  {journal} {Phys. Rev. B}\ }\textbf {\bibinfo {volume} {88}},\
  \bibinfo {pages} {241105} (\bibinfo {year} {2013})}\BibitemShut {NoStop}%
\bibitem [{\citenamefont {Haldane}(1983{\natexlab{b}})}]{Haldane1983}%
  \BibitemOpen
  \bibfield  {author} {\bibinfo {author} {\bibfnamefont {F.~D.~M.}\
  \bibnamefont {Haldane}},\ }\href {\doibase 10.1103/PhysRevLett.51.605}
  {\bibfield  {journal} {\bibinfo  {journal} {Phys. Rev. Lett.}\ }\textbf
  {\bibinfo {volume} {51}},\ \bibinfo {pages} {605} (\bibinfo {year}
  {1983}{\natexlab{b}})}\BibitemShut {NoStop}%
\bibitem [{\citenamefont {Jain}(1989)}]{Jain1989}%
  \BibitemOpen
  \bibfield  {author} {\bibinfo {author} {\bibfnamefont {J.~K.}\ \bibnamefont
  {Jain}},\ }\href {\doibase 10.1103/PhysRevLett.63.199} {\bibfield  {journal}
  {\bibinfo  {journal} {Phys. Rev. Lett.}\ }\textbf {\bibinfo {volume} {63}},\
  \bibinfo {pages} {199} (\bibinfo {year} {1989})}\BibitemShut {NoStop}%
\bibitem [{\citenamefont {Jain}(2007)}]{JainBook}%
  \BibitemOpen
  \bibfield  {author} {\bibinfo {author} {\bibfnamefont {J.~K.}\ \bibnamefont
  {Jain}},\ }\href@noop {} {\emph {\bibinfo {title} {Composite Fermions}}}\
  (\bibinfo  {publisher} {Cambridge University Press},\ \bibinfo {year}
  {2007})\BibitemShut {NoStop}%
\bibitem [{\citenamefont {Haldane}\ and\ \citenamefont
  {Shen}(2016)}]{Haldane2016}%
  \BibitemOpen
  \bibfield  {author} {\bibinfo {author} {\bibfnamefont {F.~D.~M.}\
  \bibnamefont {Haldane}}\ and\ \bibinfo {author} {\bibfnamefont
  {Y.}~\bibnamefont {Shen}},\ }\href@noop {} {\bibfield  {journal} {\bibinfo
  {journal} {arXiv:cond-mat}\ ,\ \bibinfo {pages} {1512.04502}} (\bibinfo
  {year} {2016})}\BibitemShut {NoStop}%
\end{thebibliography}%

\end{document}